\documentclass[prl,aps,showpacs,twocolumn,superscriptaddress,floatfix,nofootinbib]{revtex4-1}
\usepackage[utf8]{inputenc}
\usepackage{bm}
\usepackage{bbold}
\usepackage{mathtools}
\usepackage[toc,page]{appendix}
\usepackage{dsfont}
\usepackage{combelow}
\usepackage[colorlinks=true,linkcolor=blue,citecolor=blue, urlcolor=blue]{hyperref} 
\usepackage[a4paper,top=2.4cm,bottom=2.4cm,right=1.0cm,bindingoffset=-1.2cm]{geometry}
\usepackage{tikz}
\usepackage{graphicx}
\usepackage{subfigure}
\usepackage{scalerel,amssymb}
\DeclareMathAlphabet{\pazocal}{OMS}{zplm}{m}{n}
\usepackage{ulem}

\renewcommand{\d}{\mathrm{d}}

\newcommand{\xT}{{\mathbf{x}_\perp}}

\newcommand{\eavg}[1]{\left\langle #1 \right\rangle_\epsilon}

\begin{document}
\title{Establishing the Range of Applicability of Hydrodynamics in High-Energy Collisions }
\author{Victor E. Ambru\cb{s}}
\affiliation{Department of Physics, West University of Timi\cb{s}oara, \\
Bd.~Vasile P\^arvan 4, Timi\cb{s}oara 300223, Romania}
\author{S.~Schlichting}
\affiliation{Fakultät für Physik, Universität Bielefeld, D-33615 Bielefeld, Germany}
\author{C.~Werthmann}
\email{cwerthmann@physik.uni-bielefeld.de}
\affiliation{Fakultät für Physik, Universität Bielefeld, D-33615 Bielefeld, Germany}
\affiliation{Incubator of Scientific Excellence-Centre for Simulations of Superdense Fluids, University of Wrocław, pl. Maxa Borna 9, 50-204 Wrocław, Poland}
\date{\today}

\begin{abstract}
We simulate the space-time dynamics of high-energy collisions based on a microscopic kinetic description in the conformal relaxation time approximation, in order to determine the range of applicability of an effective description in relativistic viscous hydrodynamics. We find that hydrodynamics provides a quantitatively accurate description of collective flow when  the average inverse Reynolds number $\text{Re}^{-1}$ is sufficiently small and the early pre-equilibrium stage is properly accounted for. We further discuss the implications of our findings for the (in)applicability of hydrodynamics in proton-proton, proton-nucleus and light nucleus collisions.
\end{abstract}

\maketitle

\textit{Introduction.}
Relativistic heavy ion collisions probe the behaviour of strong interaction matter under extreme conditions. One of the central goals of these experiments is to determine the properties of the Quark-Gluon-Plasma (QGP), a new phase of deconfined strongly interacting matter. The emergence of collective phenomena in such collisions has been successfully described using relativistic viscous hydrodynamics~\cite{Teaney:2009qa,Song:2010mg,Gale:2013da,Heinz:2013th,Luzum:2013yya,Jeon:2015dfa}, which in modern simulation frameworks forms the central part of multistage evolution models~\cite{Putschke:2019yrg,JETSCAPE:2020mzn,Nijs:2020roc}.

Despite tremendous phenomenological success, it is important to remember that viscous relativistic hydrodynamics is by construction an effective macroscopic description of the underlying microscopic theory of quantum chromodynamics (QCD), which aims to describe the long-time and  long-wavelength behaviour of QCD close to equilibrium.
Consequently, the applicability of hydrodynamics requires a separation of the timescale and length scale of the dynamics of the microscopic constituents and those of the macroscopic dynamics of the system as a whole
as well as some degree of equilibration of the QGP, both of which are not necessarily fulfilled in heavy ion collisions. While for sufficiently high multiplicity events one can expect the system to quickly evolve toward equilibrium, establishing the timescale of this process for practical purposes 
%so far has mostly been 
has so far been mostly
guesswork~\cite{Heinz:2013th}.
In recent years, the observation of collective flow phenomena even in small collision systems, that were traditionally considered to be too dilute to allow for QGP formation, has challenged the hydrodynamic paradigm~\cite{ALICE:2014dwt,ATLAS:2017hap,CMS:2017kcs,Dusling:2015gta,Loizides:2016tew,Nagle:2018nvi}. While remarkable progress has been made in understanding the emergence and applicability of hydrodynamic behavior in the simplistic $0+1D$ Bjorken flow~\cite{Berges:2013fga,Berges:2013lsa,Heller:2015dha,Spalinski:2017mel,Strickland:2017kux,Strickland:2018ayk,Spalinski:2018mqg,Giacalone:2019ldn,Kurkela:2019set,Denicol:2019lio,Almaalol:2020rnu,Heller:2020anv,Du:2020zqg,Blaizot:2021cdv,Chattopadhyay:2021ive,Du:2022bel}, despite some notable attempts~\cite{Denicol:2014tha,Kurkela:2019kip,Kurkela:2020wwb} the crucial question under what circumstances viscous hydrodynamics provides a reliable and accurate description of the more complex space-time dynamics of real-world collisions remains largely unanswered.

In this Letter we employ a microscopic description in relativistic kinetic theory to determine the range of applicability of viscous hydrodynamics in high-energy collisions. Starting from a non-equilibrium initial state immediately after the collision, we find that -- even in the limit of very large interaction strength -- the early time pre-equilibrium dynamics can never be described by ordinary viscous hydrodynamics which can affect collective flow observables at the few-percent level. Subsequently, for sufficiently large systems, the fluid approaches equilibrium before the onset of the transverse expansion and the development of (anisotropic) transverse flow can nevertheless be described macroscopically using viscous relativistic hydrodynamics. By matching the non-equilibrium kinetic description to relativistic viscous hydrodynamics, we determine a critical inverse Reynolds number $\text{Re}^{-1}_{c}$ below which hydrodynamics can describe the space-time dynamics of the QGP, including the development of anisotropic transverse flow, with a few percent accuracy. 

Conversely, for small systems or large viscosity, the system remains out-of-equilibrium over the course of the evolution and we evaluate when the conditions for the applicability of viscous hydrodynamics are met as a function of shear viscosity to entropy density ratio $\eta/s$, initial state energy density and system size, and thereby infer bounds on the applicability of hydrodynamics in small systems.

Since the reasons for the failure of hydrodynamics at early times can be understood and cured within effectively $0+1D$ Bjorken flow dynamics, we further propose an alternative scheme that allows us to initialize hydrodynamic simulations immediately after the collision $\tau \to 0$ by compensating for the improper description at early times through rescaling the initial conditions. 

In this Letter, we focus on the phenomenologically relevant findings and insights noting that our companion paper~\cite{Ambrus:2022koq} contains a variety of additional information and details related to our study.

\textit{Kinetic theory setup.}
In order to study the microscopic evolution of the space-time dynamics, we employ the Boltzmann equation in the relaxation time approximation (RTA),
\begin{eqnarray}
p^{\mu}\partial_{\mu} f(x,p)= -\frac{u^{\mu}(x)p_{\mu}}{\tau_R(x)} \left[f(x,p)-f_{\rm eq}(x,p)\right],
\end{eqnarray}
with a conformal relaxation time $\tau_R(x)=\frac{\eta/s}{5T(x)}$.
The equilibrium distribution $f_{\rm eq}$ is determined by the temperature $T(x)$ and flow velocity $u^{\mu}(x)$, which are 
obtained via Landau matching 
$u_{\mu}T^{\mu\nu}=\epsilon u^{\nu}$, where $\epsilon=a T^4$ denotes the energy density, $a=\nu_{\rm eff}\pi^2/30 $ 
 with $\nu_{\rm eff}$ being the effective number of (bosonic) degrees of freedom and $T^{\mu\nu}(x)=\int_{p}p^{\mu}p^{\nu} f(x,p)$ 
 is the energy-momentum tensor (see~\cite{Ambrus:2022koq} for details). Because of the particular simplicity of this microscopic theory, differences in the space-time dynamics of the collision for a fixed initial energy density profile $(\epsilon\tau)_{0}(\xT)$ only depend on a single dimensionless opacity parameter~\cite{Kurkela:2019kip,Ambrus:2021fej}
\begin{eqnarray}
\hat{\gamma}=\frac{1}{5 \eta/ s} 
 \left(\frac{R}{\pi a} \frac{\d E_\perp^0}{\d\eta}\right)^{1/4},
 \label{eq:ghat}
\end{eqnarray}
which accounts for variations of 
the shear viscosity to entropy density ratio 
$\eta/s$, as well as of the initial energy per unit rapidity $dE^{0}_{\bot}/d\eta=\int_{\xT} (\epsilon\tau)_0(\xT)$ and the system size $R^2=(dE^{0}_{\bot}/d\eta)^{-1} \int_{\xT} {\bf x}_{\bot}^{2} (\epsilon\tau)_0(\xT)$. In order to have a well-defined collision geometry, we focus on $\mathrm{Pb}+\mathrm{Pb}$ collision at LHC energies and employ an average initial-state energy density profile $(\epsilon\tau)_{0}(\xT)$ obtained from a saturation model~\cite{Borghini:2022iym} and vary  $\eta/s$ as well as the centrality of the collision to assess the opacity dependence.

We will quantify the development of anisotropic transverse flow in terms of the elliptic energy flow,
\begin{eqnarray}
\varepsilon_{p}
=\frac{\int_{\xT} T_{xx}(\xT) -T_{yy}(\xT) +2i T_{xy}(\xT)}{\int_{\xT} T_{xx}(\xT) +T_{yy}(\xT)}, \label{eq:epsp_def}
\end{eqnarray} 
which can be inferred directly from the energy-momentum tensor $T^{\mu\nu}$, and thus eliminates uncertainties related to the hadronization process~\cite{Kapusta:1980zz,Petersen:2009mz,Huovinen:2012is}. Microscopic simulations in kinetic theory will be contrasted with the macroscopic description in relativistic hydrodynamics by employing the vHLLE Hydro code~\cite{Karpenko:2013wva}, which provides the evolution in Mueller-Israel-Stewart type second-order viscous hydrodynamics~\cite{Muller.1967,Israel.1979} with conformal equation of state and transport coefficients matched to the RTA Boltzmann equation (see~\cite{Ambrus:2022koq} for details).

\begin{figure*}[t]
    \centering
    \includegraphics[height=200pt]{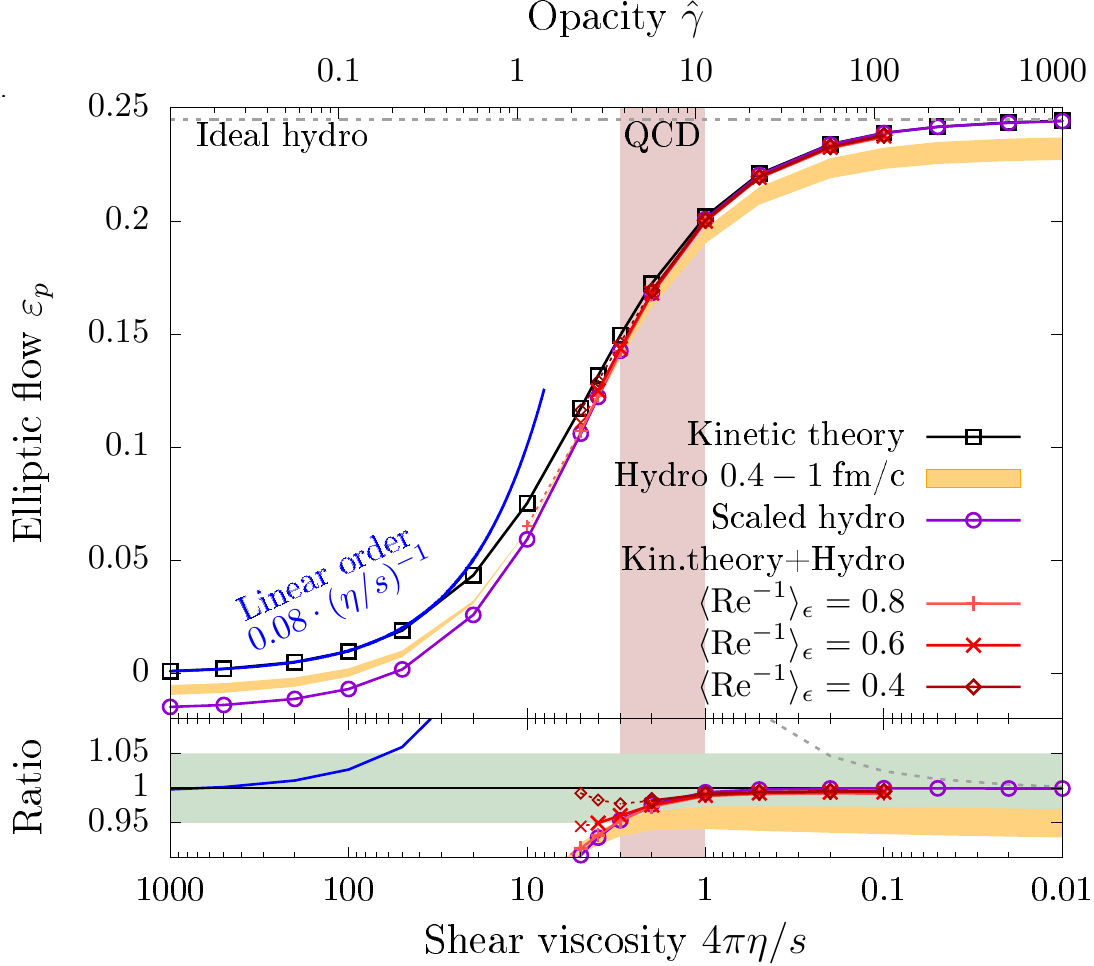}
    \includegraphics[height=200pt]{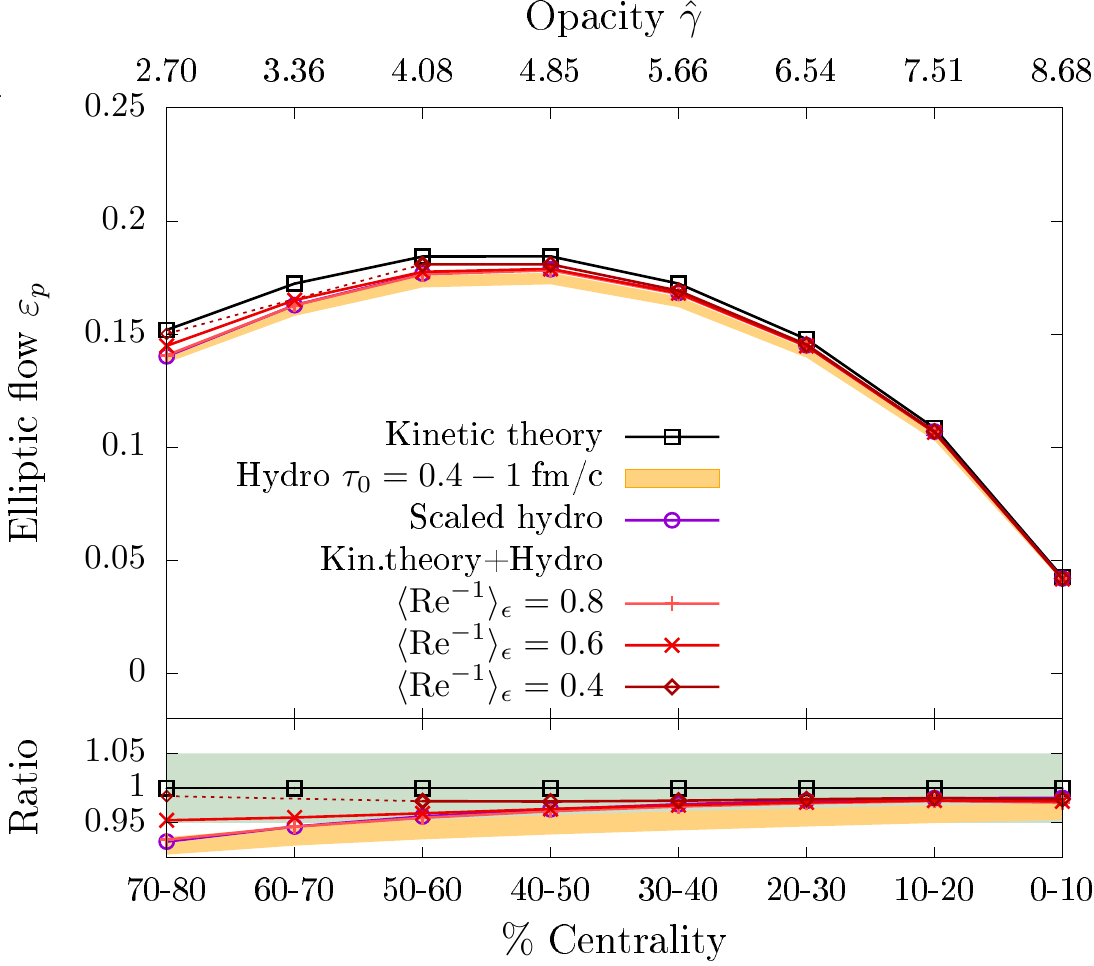}
    \caption{
    Variations of the elliptic flow (left) as a function of the shear viscosity to entropy density ratio $\eta/s$ for $30-40\%$ $\mathrm{Pb}+\mathrm{Pb}$ collisions and (right) as a function of collision centrality for fixed $\eta/s=2/4\pi$. Simulations in kinetic theory (black squares) are compared to ideal (gray dashed) and viscous hydrodynamics (purple circles) as well as hybrid simulations (red pluses, crosses and diamonds) matching kinetic theory to hydrodynamics at different values of the average inverse Reynolds number $\text{Re}^{-1}$. Bands also show the results for naive hydrodynamics simulations with varying initialization times $\tau_{0}=0.4-1.0\ {\rm fm}/c$. Semi-analytic results from a leading order opacity expansion are shown as a blue curve in the left panel.
    }
    \label{fig:master_e_p}
\end{figure*}

{\it Collective flow and applicability of hydrodynamics.}
In order to assess the range of applicability of the hydrodynamic description, we compare the results to microscopic simulations in kinetic theory. In Fig.~\ref{fig:master_e_p} we present the final results\footnote{In this letter we only present final state observables. Further discussion of the time evolution can be found in our companion paper~\cite{Ambrus:2022koq}.}
(at $\tau =4R$) for the elliptic energy flow $\varepsilon_p$ in $5.02\,$TeV $\mathrm{Pb}+\mathrm{Pb}$ collisions as a function of $\eta/s$ in mid-central collisions ($30-40\%$) in the left panel and for a realistic value of $\eta/s=2/4\pi$ as a function of centrality in the right panel. Starting with the results in kinetic theory, one immediately observes a significant opacity dependence of the final state momentum response to the initial state geometry. In the limit of low opacity ($\eta/s \gg 1$), the anisotropic flow is generated by rare final state interactions and can be well described by the leading order opacity expansion~\cite{Heiselberg:1998es,Borghini:2010hy,Romatschke:2018wgi,Kurkela:2018ygx,Borghini:2018xum,Kurkela:2021ctp,Bachmann:2022cls} up to $\hat{\gamma} \lesssim 1$, as indicated by the blue line. Subsequently, for smaller values of $\eta/s$ the anisotropic flow response increases monotonically as a function of opacity and eventually saturates at large opacities ($\eta/s \ll 1$); it is further interesting to observe that the anticipated values for QCD $4\pi \eta/s \simeq 1 - 3$~\cite{Bernhard:2019bmu,JETSCAPE:2020mzn,Nijs:2020roc}  fall into a regime where the final state response exhibits a significant opacity dependence, where changes of the viscosity $\eta/s$ by fifty percent result in changes of the anisotropic flow of about $15\%$.

Now that we have established the baseline from kinetic theory, we can compare the microscopic results to the ones obtained using a hydrodynamic description. Clearly, the first thing to note is that in the limit of infinite opacities ($\eta/s \to 0$), the kinetic theory results converge toward ideal hydrodynamics as indicated by the horizontal line. However, this seemingly obvious agreement is in fact rather non-trivial as we shall explain now. Because of the rapid longitudinal expansion, the system is initially unable to sustain a significant longitudinal pressure, which is only built up over the course of the thermalization process on a timescale $\tau_{\rm eq} \sim \hat{\gamma}^{-4/3}$~\cite{Ambrus:2021fej}. Based on the conformal behavior of the system, this isotropization of the pressure proceeds more rapidly in the hotter regions than in the colder regions of the plasma. Since the system performs work against the longitudinal expansion~\cite{Bjorken:1982qr,Giacalone:2019ldn}, the pre-equilibrium evolution of the longitudinal pressure affects the evolution of the energy density, resulting in the phenomenon of inhomogeneous longitudinal cooling~\cite{Ambrus:2021fej}, where hotter regions of the plasma begin to cool faster than colder regions of the plasma, thus leading to a small but non-negligible change of the geometry of the energy density profile even prior to the onset of the (anisotropic) transverse expansion. In contrast, in ideal hydrodynamics the system is always assumed to be in an isotropic local thermal equilibrium state, and the effect of inhomogeneous longitudinal cooling is absent. Hence, in order to restore agreement with ideal hydrodynamics in the limit of infinite opacities ($\eta/s \to 0)$, it is in fact necessary to initialize the ideal hydrodynamic simulation with the equilibrated energy density profile (as opposed to the original profile in the limit $\tau \to0$) as is done in Fig.~\ref{fig:master_e_p}.

While inhomogeneous longitudinal cooling is absent in ideal hydrodynamics, viscous hydrodynamics describes this effect incorrectly as it generically features a negative longitudinal pressure at very early times~\cite{Kurkela:2019set}. Similar to the case of ideal hydrodynamics, this effect can be compensated by a local rescaling of the initial energy density profile.
Specifically, we demand that under the $0+1$-D Bjorken flow evolution, the late-time behaviour of the energy density agrees between hydrodynamics and kinetic theory.
Mathematically, we make use of the attractor solution for the non-equilibrium evolution of the energy density~\cite{Giacalone:2019ldn,Jankowski:2020itt}
\begin{equation}
\tau^{4/3}\epsilon = \frac{\tau_0^{4/3} \epsilon_0 }{\mathcal{E}(\tilde{w}_0)} \mathcal{E}(\tilde{w}),
\label{eq:attractor}
\end{equation}
where $\mathcal{E}$ is a universal function that depends only on the conformal scaling variable $\tilde{w} = \tau T / (4\pi \eta / s)$, with its asymptotic behaviour at early and late times given by
\begin{equation}
 \mathcal{E}(\tilde{w} \ll 1) = C_\infty^{-1} \tilde{w}^\gamma, \qquad
 \mathcal{E}(\tilde{w} \gg 1) = 1 - 
 \frac{1}{4\pi \tilde{w}}.
 \label{eq:limits}
\end{equation}
Crucially, the coefficients $C_\infty$ and $\gamma$ that describe the longitudinal cooling at early times take different values in kinetic theory $(\gamma=4/9\;,~C_\infty^{\rm RTA} \simeq 0.88)$ and viscous hydrodynamics $(\gamma = (\sqrt{505}-13)/18 \simeq 0.526\;,~C_\infty^{\rm Hydro}\simeq 0.82)$, whereas in ideal hydrodynamics they are trivially determined as $\mathcal{E}(\tilde{w}) = 1$, $C_\infty = 1$ and $\gamma = 0$. By taking advantage of the fact that in kinetic theory $(\epsilon\tau)_{0}(\xT)$ is constant at early times, we then initialize the energy density in hydrodynamics as
 \begin{eqnarray}
 \label{eq:scaling}
  &&\epsilon^{\rm Hydro}_{0}(\xT) = \\ 
  &&\quad \left[ \left(\frac{4\pi\eta/s}{\tau_0} a^{\frac{1}{4}} 
 \right)^{\frac{1}{2} - \frac{9\gamma}{8}} 
 \left(\frac{C_\infty^{\rm RTA}}{C_\infty^{\rm Hydro}}\right)^{9/8}
 \frac{(\epsilon\tau)_{0}(\xT)}{\tau_0}\right]^{\frac{8/9}{1 - \gamma/4}} \nonumber
\end{eqnarray}
such that upon substituting Eq.~\eqref{eq:scaling} into Eq.~\eqref{eq:attractor}, hydrodynamics and kinetic theory agree at the level of $\tau^{4/3} \epsilon$ when $\tilde{w} \gg 1$ and $\tilde{w}_0 \ll 1$, by virtue of Eq.~\eqref{eq:limits}.

\begin{figure*}[t]
    \centering
    \includegraphics[height=200pt]{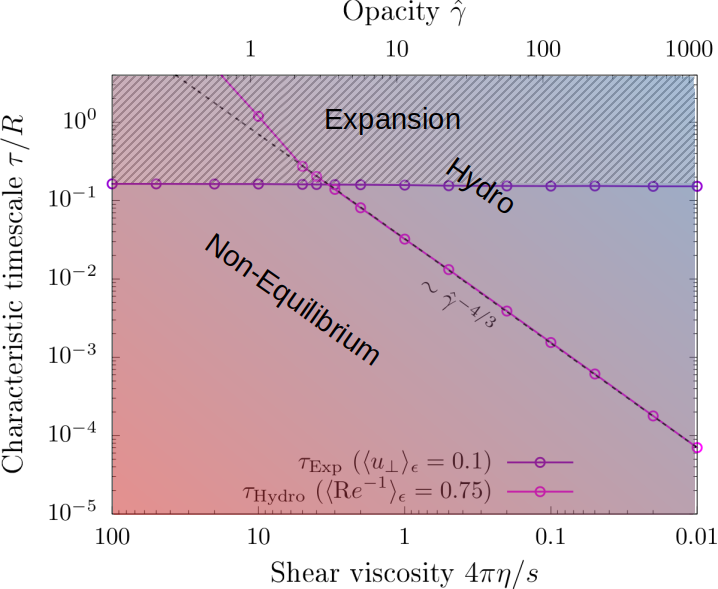}
    \includegraphics[height=200pt]{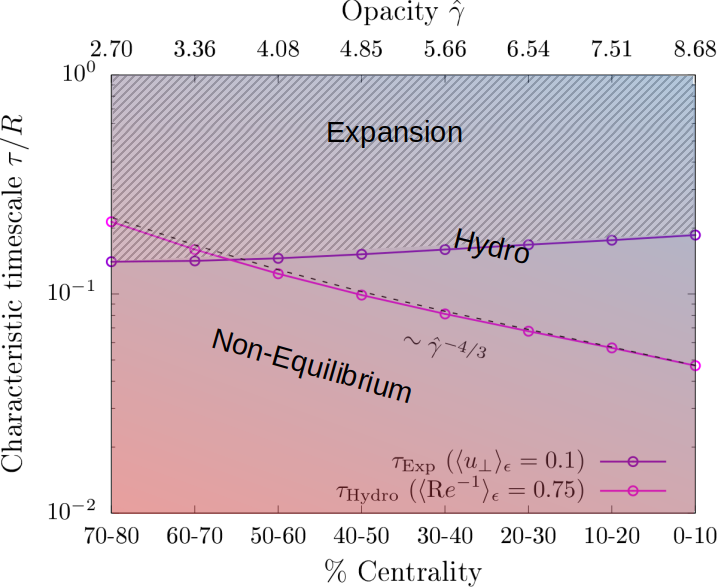}
    \caption{Characteristic timescales for the onset of the transverse expansion $\tau_{\rm Exp}$ and the onset of hydrodynamic behavior $\tau_{\rm Hydro}$ (left) as a function of 
    $\eta/s$ for $30-40\%$ $\mathrm{Pb}+\mathrm{Pb}$ collisions and (right) as a function of collision centrality for fixed $4\pi\eta/s=2$. Dashed lines indicate the $\hat{\gamma}^{-4/3}$ estimate for the transition between non-equilibrium and hydrodynamic behavior in Eq.~(\ref{eq:tHydroParam}). Below opacities $\hat{\gamma} \simeq 3-4$, the transverse expansion sets in while the system is significantly out of equilibrium and hydrodynamics is unable to accurately describe the transverse expansion of the system.
    }
    \label{fig:tc_per_opacity}
\end{figure*}

By performing such a local rescaling, viscous hydrodynamics can be initialized at arbitrarily early times ($\tau_{0} \to 0)$ and provides an accurate description of the anisotropic flow down to opacities of $\hat{\gamma} \gtrsim 3$, as can be seen from the purple curve in Fig.~\ref{fig:master_e_p}. We finally note that, if the pre-equilibrium regime is completely ignored and hydrodynamic simulations are naively initialized with the original energy density profile at a fixed proper time $\tau_0=0.4 - 1\, {\rm fm}/c$, the above effects and the absence of pre-flow lead to sizeable deviations even in the limit of very large opacities. In all cases, similar conclusions can also be reached for the development of radial flow and the cooling of the plasma, as demonstrated explicitly in the Supplemental Material (see below).

{\it Critical inverse Reynolds number.}
Because of the aforementioned subtleties associated with the pre-equilibrium stage, another viable alternative is to employ a kinetic description at early times when the system is far from equilibrium and switch to a macroscopic description only once the system is sufficiently close to local thermal equilibrium for hydrodynamics to be applicable. While in phenomenological studies the timescale for initializing hydrodynamic simulations is typically chosen as $\sim 1\, {\rm fm}/c$~\cite{Heinz:2013th}, a more physical choice can be achieved by monitoring the magnitude of non-equilibrium corrections, as characterized by the (average) inverse Reynolds number~\cite{Niemi:2014wta}\footnote{Our definition follows the convention of Ref.~\cite{Denicol:2012cn}, by which ${\rm Re}^{-1}$ is the ratio between dissipative and equilibrium terms. Note that this differs from the convention commonly used in non-relativistic fluid dynamics (also in Ref.~\cite{Floerchinger:2011pxy} for heavy ion collisions), where ${\rm Re}^{-1}{=} \eta / \rho u L$ is the ratio of the kinematic viscosity $\eta / \rho$ to the typical velocity and length scales of the system. Our normalisation factor is conventional and chosen such that ${\rm Re}^{-1}{=}1$ in the limit $\tau{\to}0$, where the energy-momentum tensor in kinetic theory assumes the form $T^{\mu\nu}{=}\text{diag}(\epsilon,\epsilon/2,\epsilon/2,0)$.}
\begin{eqnarray}
{\rm Re}^{-1} = \left(\frac{6 \pi^{\mu\nu} \pi_{\mu\nu}}{\epsilon^2}\right)^{1/2},\label{eq:Reinv_def}
\end{eqnarray}
where the shear-stress tensor $\pi^{\mu\nu}$ is the non-equilibrium part of $T^{\mu\nu}$.
By switching from kinetic theory to hydrodynamics at different values of
$\text{Re}^{-1}$, we can then infer what degree of equilibration is
required for hydrodynamics to provide an accurate description of the
space-time dynamics.

Simulation results for hybrid simulations using kinetic theory +
hydrodynamics are also presented in Fig.~\ref{fig:master_e_p} and are in excellent
agreement with the microscopic calculations from kinetic theory at large
opacities $\hat{\gamma} \gg 1$. Strikingly, when comparing the results
obtained by switching at different $\text{Re}^{-1}=0.8,0.6,0.4$ one observes that the
agreement between kinetic theory and hybrid simulations improves
significantly with decreasing the inverse Reynolds number at the
point of switching. Specifically, when switching at ${\rm Re}^{-1} \lesssim
0.75$, the discrepancy between kinetic theory and hybrid simulations
remains below $5\%$ irrespective of the viscosity or centrality of the
collision. We note, however, that for large viscosities ($4\pi \eta/s
\gtrsim 4$) as well as for more peripheral collisions ($\gtrsim 60\%$), a
sufficiently small inverse Reynolds number is only achieved after a long period of evolution in kinetic theory; in order
to distinguish this behavior, curves for which the switching time
$\tau/R$ exceeds the value $0.5$ are shown with dashed lines in
Fig.~\ref{fig:master_e_p}.
 Eventually, for very large viscosities $4\pi\eta/s\gtrsim 10$ or very peripheral collisions $\gtrsim 80\%$, the desired value of $\text{Re}^{-1}$ may never be reached throughout the evolution of the system.

\textit{Range of applicability of hydrodynamics.} Based on the observation that a
critical inverse Reynolds number $\text{Re}^{-1}_{c} \approx 0.75$ is required
for the applicability of viscous hydrodynamics, we can immediately rule
out the viability of a hydrodynamic description at 
small opacities $\hat{\gamma} \lesssim 1$, where the system remains significantly
out-of-equilibrium throughout its entire evolution and this threshold is
never reached. However at smaller viscosities, the aforementioned
threshold may be reached very early or at a comparatively late time and
the question whether or not hydrodynamics is applicable becomes a more
delicate issue. Hence, in order to quantify whether or not hydrodynamics
provides a meaningful and accurate description of the space-time
dynamics of high-energy collisions, we will compare the timescale $\tau_{\rm Hydro}$ for
the equilibration of the system, determined by requiring $\text{Re}^{-1} <
\text{Re}^{-1}_{c}$, with the onset of the transverse expansion $\tau_{\rm Exp}$,
determined by requiring that the average transverse flow velocity
$\langle u_\perp\rangle_{\epsilon}$ becomes $0.1$ times the speed of light, as depicted in Fig.~\ref{fig:tc_per_opacity}.

Irrespective of the opacity, the transverse expansion sets in on timescales $\tau_{\rm Exp}/R\approx 0.2$ albeit with a slight dependence on the initial energy
density profile, as can be seen from the centrality dependence of the
curve in the right panel. Conversely, the timescale for the applicability of hydrodynamics shows a strong opacity dependence~\cite{Ambrus:2021fej}, which can be quantified semi-empirically as\footnote{While the factor $\hat{\gamma}^{-4/3}$ and the proportionality to $(\text{Re}^{-1}_{c})^{-3/2}$ for $(\text{Re}^{-1}_{c}) \ll 1$ follow from the properties of the Bjorken flow attractor, the numerical pre-factor and the correction term have been determined from a fit to our simulation results.}
\begin{equation}
\label{eq:tHydroParam}
\tau_{\rm Hydro}/R \approx 1.53~\hat{\gamma}^{-4/3}~\left[(\text{Re}^{-1}_{c})^{-3/2}-1.21 (\text{Re}^{-1}_{c})^{0.7}\right]\;, %\nonumber
\end{equation}
as indicated by the dashed lines in Fig.~\ref{fig:tc_per_opacity}. By comparing the different curves in Fig.~\ref{fig:tc_per_opacity}, one then concludes that for opacities
\begin{eqnarray}
\hat{\gamma} \gtrsim 3-4,
\end{eqnarray} 
which in $\mathrm{Pb}+\mathrm{Pb}$ collisions corresponds to $\eta/s \lesssim 3/4\pi$ for $30-40\%$ centrality or 
$\lesssim 60\%$ centrality for $\eta/s=2/4\pi$, the system undergoes equilibration (well) before the onset
of the transverse expansion ($\tau_{\rm Hydro}<\tau_{\rm Exp}$), such that viscous hydrodynamics provides a meaningful and accurate description of the development
of (anisotropic) transverse flow. Conversely, for $\hat{\gamma} \lesssim 3-4$ hydrodynamics is not applicable as the system
remains out-of-equilibrium during the transverse expansion ($\tau_{\rm Hydro}>\tau_{\rm Exp}$) and a genuine non-equilibrium description is required instead. \\

\textit{Conclusions.} Based on a detailed comparison with the microscopic evolution in kinetic
theory, we find that viscous hydrodynamics provides an accurate
description of the space time dynamics of high-energy heavy-ion collisions, if and only if the system is
sufficiently close to equilibrium, as quantified by an inverse Reynolds
number (cf. Eq.~\eqref{eq:Reinv_def}) $\text{Re}^{-1} \lesssim 0.75$. Clearly, this is
not the case during the very early stages of the collision, where
irrespective of the shear viscosity to entropy density ratio $(\eta/s)$ the system
is highly anisotropic and the inhomogeneous longitudinal cooling cannot
be properly described within ordinary viscous hydrodynamics.
Disregarding this effect leads to percent-level deviations in the
development of anisotropic transverse flow even in the limit of very
small viscosities. However, for sufficiently small viscosities, this
effect can be compensated by an inhomogeneous local rescaling of the
initial energy density profile (cf. Eq.~\eqref{eq:scaling}), or by employing a hybrid description
where the energy momentum tensor after the early kinetic evolution 
provides the initial conditions for the subsequent hydrodynamic
stage. It is also conceivable that improved hydrodynamic theories, such
as anisotropic hydrodynamics~\cite{Martinez:2010sc,Florkowski:2010cf,Florkowski:2013lya,Martinez:2012tu,McNelis:2021zji}, or hybrid schemes based on the core-corona picture~\cite{Werner:2007bf,Kanakubo:2019ogh} can further push the limits of applicability by improving the
description far from equilibrium, and it will be interesting to investigate 
this further in the future.

Similarly, in very peripheral collisions or if the shear viscosity of the QGP was significantly
 larger ($\eta/s \gtrsim 3/4\pi$), the system remains out-of-equilibrium for a significant part 
 or even all of its space-time evolution, such that, if at all, hydrodynamics only becomes applicable
at very late times, when the transverse expansion is already very significant. Based on our analysis, we argued
that a meaningful and accurate hydrodynamic description of collective flow can be achieved 
if the opacity $\hat{\gamma}$ (c.f. Eq.~\eqref{eq:ghat}) exceeds values of $\approx 3-4$.

Since within our simplified kinetic description, the non-equilibrium evolution is governed by the single dimensionless opacity 
parameter $\hat{\gamma}$, it is tempting to speculate about the consequences of our findings for the theoretical description
of small collision systems. Based on experimental measurements of the transverse energy per unit rapidity $dE_{\bot}/d\eta$
and theoretical estimates of the system size, both of which are detailed in the Supplemental Material (see below), we can deduce 
that for a typical value of $\eta/s=2/4\pi$ hydrodynamics is not applicable in minimum bias $p+p$ collisions at LHC energies where opacities typically take values of
$\hat{\gamma}\approx 0.7$. Interestingly the situation is more subtle in $p+\mathrm{Pb}$ collisions where opacities range from
$\hat{\gamma}\approx 1.5$ in minimum bias events toward the limits of applicability of hydrodynamics in high-multiplicity events
 where $\hat{\gamma}\approx 2.7$ can be reached. Unfortunately though, the initial state geometry of $p+\mathrm{Pb}$ collisions is at present also poorly constrained~\cite{Schenke:2014zha,Mantysaari:2022ffw,Demirci:2022wuy}, such that it becomes difficult to disentangle the effects of the geometry from the effects of the flow response. 
 However, the onset of hydrodynamic behavior will also be explored in $\mathrm{O}+\mathrm{O}$ collisions where at LHC energies the estimated opacities reach from $1.4$ in peripheral events and $2.2$ in mid-central events all the way to $3.1$ in central events. 
 Since, as advertised in~\cite{Brewer:2021kiv}, the collision geometry in $\mathrm{O}+\mathrm{O}$ collisions is much better constrained, we therefore expect that
 such collisions will not only provide a crucial test of the applicability of hydrodynamics in heavy-ion collisions but also
 penetrate into the exciting regime of non-equilibrium QCD in mid-central and peripheral events.

\begin{acknowledgments}
\textit{Acknowledgements:} 
We thank P.~Aasha, N.~Borghini, H.~Elfner, A.~Mazeliauskas,  H.~Roch, A.~Shark, and U.~A.~Wiedemann for valuable discussions. We also thank S. Busuioc for carefully reading the manuscript. This work is supported by the Deutsche Forschungsgemeinschaft (DFG, German Research Foundation)
through the CRC-TR 211 ’Strong-interaction matter under extreme conditions’– project
number 315477589 – TRR 211. V.E.A.~gratefully acknowledges the support through a grant of the 
Ministry of Research, Innovation and Digitization, CNCS - UEFISCDI,
project number PN-III-P1-1.1-TE-2021-1707, within PNCDI III. C.W. was supported by the program Excellence Initiative–Research University of the University of Wrocław of the Ministry of Education and Science. Numerical calculations presented in this work were performed at the Paderborn Center for Parallel Computing (PC2) and the Center for Scientific Computing (CSC) at the Goethe-University of Frankfurt and we gratefully acknowledge their support. 
\end{acknowledgments}

\bibliography{references.bib}

\begin{thebibliography}{70}%
\makeatletter
\providecommand \@ifxundefined [1]{%
 \@ifx{#1\undefined}
}%
\providecommand \@ifnum [1]{%
 \ifnum #1\expandafter \@firstoftwo
 \else \expandafter \@secondoftwo
 \fi
}%
\providecommand \@ifx [1]{%
 \ifx #1\expandafter \@firstoftwo
 \else \expandafter \@secondoftwo
 \fi
}%
\providecommand \natexlab [1]{#1}%
\providecommand \enquote  [1]{``#1''}%
\providecommand \bibnamefont  [1]{#1}%
\providecommand \bibfnamefont [1]{#1}%
\providecommand \citenamefont [1]{#1}%
\providecommand \href@noop [0]{\@secondoftwo}%
\providecommand \href [0]{\begingroup \@sanitize@url \@href}%
\providecommand \@href[1]{\@@startlink{#1}\@@href}%
\providecommand \@@href[1]{\endgroup#1\@@endlink}%
\providecommand \@sanitize@url [0]{\catcode `\\12\catcode `\$12\catcode
  `\&12\catcode `\#12\catcode `\^12\catcode `\_12\catcode `\%12\relax}%
\providecommand \@@startlink[1]{}%
\providecommand \@@endlink[0]{}%
\providecommand \url  [0]{\begingroup\@sanitize@url \@url }%
\providecommand \@url [1]{\endgroup\@href {#1}{\urlprefix }}%
\providecommand \urlprefix  [0]{URL }%
\providecommand \Eprint [0]{\href }%
\providecommand \doibase [0]{http://dx.doi.org/}%
\providecommand \selectlanguage [0]{\@gobble}%
\providecommand \bibinfo  [0]{\@secondoftwo}%
\providecommand \bibfield  [0]{\@secondoftwo}%
\providecommand \translation [1]{[#1]}%
\providecommand \BibitemOpen [0]{}%
\providecommand \bibitemStop [0]{}%
\providecommand \bibitemNoStop [0]{.\EOS\space}%
\providecommand \EOS [0]{\spacefactor3000\relax}%
\providecommand \BibitemShut  [1]{\csname bibitem#1\endcsname}%
\let\auto@bib@innerbib\@empty
%</preamble>
\bibitem [{\citenamefont {Teaney}(2010)}]{Teaney:2009qa}%
  \BibitemOpen
  \bibfield  {author} {\bibinfo {author} {\bibfnamefont {D.~A.}\ \bibnamefont
  {Teaney}},\ }\enquote {\bibinfo {title} {{Viscous Hydrodynamics and the Quark
  Gluon Plasma}},}\ in\ \href {\doibase 10.1142/9789814293297_0004} {\emph
  {\bibinfo {booktitle} {{Quark-gluon plasma 4}}}},\ \bibinfo {editor} {edited
  by\ \bibinfo {editor} {\bibfnamefont {R.~C.}\ \bibnamefont {Hwa}}\ and\
  \bibinfo {editor} {\bibfnamefont {X.-N.}\ \bibnamefont {Wang}}}\ (\bibinfo
  {publisher} {World Scientific Publishing Co., Singapore},\ \bibinfo {year}
  {2010})\ pp.\ \bibinfo {pages} {207--266},\ \Eprint
  {http://arxiv.org/abs/0905.2433} {arXiv:0905.2433 [nucl-th]} \BibitemShut
  {NoStop}%
\bibitem [{\citenamefont {Song}\ \emph {et~al.}(2011)\citenamefont {Song},
  \citenamefont {Bass}, \citenamefont {Heinz}, \citenamefont {Hirano},\ and\
  \citenamefont {Shen}}]{Song:2010mg}%
  \BibitemOpen
  \bibfield  {author} {\bibinfo {author} {\bibfnamefont {H.}~\bibnamefont
  {Song}}, \bibinfo {author} {\bibfnamefont {S.~A.}\ \bibnamefont {Bass}},
  \bibinfo {author} {\bibfnamefont {U.}~\bibnamefont {Heinz}}, \bibinfo
  {author} {\bibfnamefont {T.}~\bibnamefont {Hirano}}, \ and\ \bibinfo {author}
  {\bibfnamefont {C.}~\bibnamefont {Shen}},\ }\href {\doibase
  10.1103/PhysRevLett.106.192301} {\bibfield  {journal} {\bibinfo  {journal}
  {Phys. Rev. Lett.}\ }\textbf {\bibinfo {volume} {106}},\ \bibinfo {pages}
  {192301} (\bibinfo {year} {2011})}\BibitemShut {NoStop}%
\bibitem [{\citenamefont {Gale}\ \emph {et~al.}(2013)\citenamefont {Gale},
  \citenamefont {Jeon},\ and\ \citenamefont {Schenke}}]{Gale:2013da}%
  \BibitemOpen
  \bibfield  {author} {\bibinfo {author} {\bibfnamefont {C.}~\bibnamefont
  {Gale}}, \bibinfo {author} {\bibfnamefont {S.}~\bibnamefont {Jeon}}, \ and\
  \bibinfo {author} {\bibfnamefont {B.}~\bibnamefont {Schenke}},\ }\href
  {\doibase 10.1142/S0217751X13400113} {\bibfield  {journal} {\bibinfo
  {journal} {Int. J. Mod. Phys. A}\ }\textbf {\bibinfo {volume} {28}},\
  \bibinfo {pages} {1340011} (\bibinfo {year} {2013})},\ \Eprint
  {http://arxiv.org/abs/1301.5893} {arXiv:1301.5893 [nucl-th]} \BibitemShut
  {NoStop}%
\bibitem [{\citenamefont {Heinz}\ and\ \citenamefont
  {Snellings}(2013)}]{Heinz:2013th}%
  \BibitemOpen
  \bibfield  {author} {\bibinfo {author} {\bibfnamefont {U.}~\bibnamefont
  {Heinz}}\ and\ \bibinfo {author} {\bibfnamefont {R.}~\bibnamefont
  {Snellings}},\ }\href {\doibase 10.1146/annurev-nucl-102212-170540}
  {\bibfield  {journal} {\bibinfo  {journal} {Ann. Rev. Nucl. Part. Sci.}\
  }\textbf {\bibinfo {volume} {63}},\ \bibinfo {pages} {123} (\bibinfo {year}
  {2013})},\ \Eprint {http://arxiv.org/abs/1301.2826} {arXiv:1301.2826
  [nucl-th]} \BibitemShut {NoStop}%
\bibitem [{\citenamefont {Luzum}\ and\ \citenamefont
  {Petersen}(2014)}]{Luzum:2013yya}%
  \BibitemOpen
  \bibfield  {author} {\bibinfo {author} {\bibfnamefont {M.}~\bibnamefont
  {Luzum}}\ and\ \bibinfo {author} {\bibfnamefont {H.}~\bibnamefont
  {Petersen}},\ }\href {\doibase 10.1088/0954-3899/41/6/063102} {\bibfield
  {journal} {\bibinfo  {journal} {J. Phys. G}\ }\textbf {\bibinfo {volume}
  {41}},\ \bibinfo {pages} {063102} (\bibinfo {year} {2014})},\ \Eprint
  {http://arxiv.org/abs/1312.5503} {arXiv:1312.5503 [nucl-th]} \BibitemShut
  {NoStop}%
\bibitem [{\citenamefont {Jeon}\ and\ \citenamefont
  {Heinz}(2015)}]{Jeon:2015dfa}%
  \BibitemOpen
  \bibfield  {author} {\bibinfo {author} {\bibfnamefont {S.}~\bibnamefont
  {Jeon}}\ and\ \bibinfo {author} {\bibfnamefont {U.}~\bibnamefont {Heinz}},\
  }\href {\doibase 10.1142/S0218301315300106} {\bibfield  {journal} {\bibinfo
  {journal} {Int. J. Mod. Phys. E}\ }\textbf {\bibinfo {volume} {24}},\
  \bibinfo {pages} {1530010} (\bibinfo {year} {2015})},\ \Eprint
  {http://arxiv.org/abs/1503.03931} {arXiv:1503.03931 [hep-ph]} \BibitemShut
  {NoStop}%
\bibitem [{\citenamefont {Putschke}\ \emph {et~al.}(2019)\citenamefont
  {Putschke} \emph {et~al.}}]{Putschke:2019yrg}%
  \BibitemOpen
  \bibfield  {author} {\bibinfo {author} {\bibfnamefont {J.~H.}\ \bibnamefont
  {Putschke}} \emph {et~al.},\ }\href@noop {} {\  (\bibinfo {year} {2019})},\
  \bibinfo {note} {arXiv:1903.07706},\ \Eprint
  {http://arxiv.org/abs/1903.07706} {arXiv:1903.07706 [nucl-th]} \BibitemShut
  {NoStop}%
\bibitem [{\citenamefont {Everett}\ \emph {et~al.}(2021)\citenamefont {Everett}
  \emph {et~al.}}]{JETSCAPE:2020mzn}%
  \BibitemOpen
  \bibfield  {author} {\bibinfo {author} {\bibfnamefont {D.}~\bibnamefont
  {Everett}} \emph {et~al.} (\bibinfo {collaboration} {JETSCAPE}),\ }\href
  {\doibase 10.1103/PhysRevC.103.054904} {\bibfield  {journal} {\bibinfo
  {journal} {Phys. Rev. C}\ }\textbf {\bibinfo {volume} {103}},\ \bibinfo
  {pages} {054904} (\bibinfo {year} {2021})},\ \Eprint
  {http://arxiv.org/abs/2011.01430} {arXiv:2011.01430 [hep-ph]} \BibitemShut
  {NoStop}%
\bibitem [{\citenamefont {Nijs}\ \emph {et~al.}(2021)\citenamefont {Nijs},
  \citenamefont {van~der Schee}, \citenamefont {G\"ursoy},\ and\ \citenamefont
  {Snellings}}]{Nijs:2020roc}%
  \BibitemOpen
  \bibfield  {author} {\bibinfo {author} {\bibfnamefont {G.}~\bibnamefont
  {Nijs}}, \bibinfo {author} {\bibfnamefont {W.}~\bibnamefont {van~der Schee}},
  \bibinfo {author} {\bibfnamefont {U.}~\bibnamefont {G\"ursoy}}, \ and\
  \bibinfo {author} {\bibfnamefont {R.}~\bibnamefont {Snellings}},\ }\href
  {\doibase 10.1103/PhysRevC.103.054909} {\bibfield  {journal} {\bibinfo
  {journal} {Phys. Rev. C}\ }\textbf {\bibinfo {volume} {103}},\ \bibinfo
  {pages} {054909} (\bibinfo {year} {2021})},\ \Eprint
  {http://arxiv.org/abs/2010.15134} {arXiv:2010.15134 [nucl-th]} \BibitemShut
  {NoStop}%
\bibitem [{\citenamefont {Abelev}\ \emph {et~al.}(2014)\citenamefont {Abelev}
  \emph {et~al.}}]{ALICE:2014dwt}%
  \BibitemOpen
  \bibfield  {author} {\bibinfo {author} {\bibfnamefont {B.~B.}\ \bibnamefont
  {Abelev}} \emph {et~al.} (\bibinfo {collaboration} {ALICE}),\ }\href
  {\doibase 10.1103/PhysRevC.90.054901} {\bibfield  {journal} {\bibinfo
  {journal} {Phys. Rev. C}\ }\textbf {\bibinfo {volume} {90}},\ \bibinfo
  {pages} {054901} (\bibinfo {year} {2014})},\ \Eprint
  {http://arxiv.org/abs/1406.2474} {arXiv:1406.2474 [nucl-ex]} \BibitemShut
  {NoStop}%
\bibitem [{\citenamefont {Aaboud}\ \emph {et~al.}(2017)\citenamefont {Aaboud}
  \emph {et~al.}}]{ATLAS:2017hap}%
  \BibitemOpen
  \bibfield  {author} {\bibinfo {author} {\bibfnamefont {M.}~\bibnamefont
  {Aaboud}} \emph {et~al.} (\bibinfo {collaboration} {ATLAS}),\ }\href
  {\doibase 10.1140/epjc/s10052-017-4988-1} {\bibfield  {journal} {\bibinfo
  {journal} {Eur. Phys. J. C}\ }\textbf {\bibinfo {volume} {77}},\ \bibinfo
  {pages} {428} (\bibinfo {year} {2017})},\ \Eprint
  {http://arxiv.org/abs/1705.04176} {arXiv:1705.04176 [hep-ex]} \BibitemShut
  {NoStop}%
\bibitem [{\citenamefont {Sirunyan}\ \emph {et~al.}(2018)\citenamefont
  {Sirunyan} \emph {et~al.}}]{CMS:2017kcs}%
  \BibitemOpen
  \bibfield  {author} {\bibinfo {author} {\bibfnamefont {A.~M.}\ \bibnamefont
  {Sirunyan}} \emph {et~al.} (\bibinfo {collaboration} {CMS}),\ }\href
  {\doibase 10.1103/PhysRevLett.120.092301} {\bibfield  {journal} {\bibinfo
  {journal} {Phys. Rev. Lett.}\ }\textbf {\bibinfo {volume} {120}},\ \bibinfo
  {pages} {092301} (\bibinfo {year} {2018})},\ \Eprint
  {http://arxiv.org/abs/1709.09189} {arXiv:1709.09189 [nucl-ex]} \BibitemShut
  {NoStop}%
\bibitem [{\citenamefont {Dusling}\ \emph {et~al.}(2016)\citenamefont
  {Dusling}, \citenamefont {Li},\ and\ \citenamefont
  {Schenke}}]{Dusling:2015gta}%
  \BibitemOpen
  \bibfield  {author} {\bibinfo {author} {\bibfnamefont {K.}~\bibnamefont
  {Dusling}}, \bibinfo {author} {\bibfnamefont {W.}~\bibnamefont {Li}}, \ and\
  \bibinfo {author} {\bibfnamefont {B.}~\bibnamefont {Schenke}},\ }\href
  {\doibase 10.1142/S0218301316300022} {\bibfield  {journal} {\bibinfo
  {journal} {Int. J. Mod. Phys. E}\ }\textbf {\bibinfo {volume} {25}},\
  \bibinfo {pages} {1630002} (\bibinfo {year} {2016})},\ \Eprint
  {http://arxiv.org/abs/1509.07939} {arXiv:1509.07939 [nucl-ex]} \BibitemShut
  {NoStop}%
\bibitem [{\citenamefont {Loizides}(2016)}]{Loizides:2016tew}%
  \BibitemOpen
  \bibfield  {author} {\bibinfo {author} {\bibfnamefont {C.}~\bibnamefont
  {Loizides}},\ }\href {\doibase 10.1016/j.nuclphysa.2016.04.022} {\bibfield
  {journal} {\bibinfo  {journal} {Nucl. Phys. A}\ }\textbf {\bibinfo {volume}
  {956}},\ \bibinfo {pages} {200} (\bibinfo {year} {2016})},\ \Eprint
  {http://arxiv.org/abs/1602.09138} {arXiv:1602.09138 [nucl-ex]} \BibitemShut
  {NoStop}%
\bibitem [{\citenamefont {Nagle}\ and\ \citenamefont
  {Zajc}(2018)}]{Nagle:2018nvi}%
  \BibitemOpen
  \bibfield  {author} {\bibinfo {author} {\bibfnamefont {J.~L.}\ \bibnamefont
  {Nagle}}\ and\ \bibinfo {author} {\bibfnamefont {W.~A.}\ \bibnamefont
  {Zajc}},\ }\href {\doibase 10.1146/annurev-nucl-101916-123209} {\bibfield
  {journal} {\bibinfo  {journal} {Ann. Rev. Nucl. Part. Sci.}\ }\textbf
  {\bibinfo {volume} {68}},\ \bibinfo {pages} {211} (\bibinfo {year} {2018})},\
  \Eprint {http://arxiv.org/abs/1801.03477} {arXiv:1801.03477 [nucl-ex]}
  \BibitemShut {NoStop}%
\bibitem [{\citenamefont {Berges}\ \emph
  {et~al.}(2014{\natexlab{a}})\citenamefont {Berges}, \citenamefont
  {Boguslavski}, \citenamefont {Schlichting},\ and\ \citenamefont
  {Venugopalan}}]{Berges:2013fga}%
  \BibitemOpen
  \bibfield  {author} {\bibinfo {author} {\bibfnamefont {J.}~\bibnamefont
  {Berges}}, \bibinfo {author} {\bibfnamefont {K.}~\bibnamefont {Boguslavski}},
  \bibinfo {author} {\bibfnamefont {S.}~\bibnamefont {Schlichting}}, \ and\
  \bibinfo {author} {\bibfnamefont {R.}~\bibnamefont {Venugopalan}},\ }\href
  {\doibase 10.1103/PhysRevD.89.114007} {\bibfield  {journal} {\bibinfo
  {journal} {Phys. Rev. D}\ }\textbf {\bibinfo {volume} {89}},\ \bibinfo
  {pages} {114007} (\bibinfo {year} {2014}{\natexlab{a}})},\ \Eprint
  {http://arxiv.org/abs/1311.3005} {arXiv:1311.3005 [hep-ph]} \BibitemShut
  {NoStop}%
\bibitem [{\citenamefont {Berges}\ \emph
  {et~al.}(2014{\natexlab{b}})\citenamefont {Berges}, \citenamefont
  {Boguslavski}, \citenamefont {Schlichting},\ and\ \citenamefont
  {Venugopalan}}]{Berges:2013lsa}%
  \BibitemOpen
  \bibfield  {author} {\bibinfo {author} {\bibfnamefont {J.}~\bibnamefont
  {Berges}}, \bibinfo {author} {\bibfnamefont {K.}~\bibnamefont {Boguslavski}},
  \bibinfo {author} {\bibfnamefont {S.}~\bibnamefont {Schlichting}}, \ and\
  \bibinfo {author} {\bibfnamefont {R.}~\bibnamefont {Venugopalan}},\ }\href
  {\doibase 10.1007/JHEP05(2014)054} {\bibfield  {journal} {\bibinfo  {journal}
  {JHEP}\ }\textbf {\bibinfo {volume} {05}},\ \bibinfo {pages} {054} (\bibinfo
  {year} {2014}{\natexlab{b}})},\ \Eprint {http://arxiv.org/abs/1312.5216}
  {arXiv:1312.5216 [hep-ph]} \BibitemShut {NoStop}%
\bibitem [{\citenamefont {Heller}\ and\ \citenamefont
  {Spalinski}(2015)}]{Heller:2015dha}%
  \BibitemOpen
  \bibfield  {author} {\bibinfo {author} {\bibfnamefont {M.~P.}\ \bibnamefont
  {Heller}}\ and\ \bibinfo {author} {\bibfnamefont {M.}~\bibnamefont
  {Spalinski}},\ }\href {\doibase 10.1103/PhysRevLett.115.072501} {\bibfield
  {journal} {\bibinfo  {journal} {Phys. Rev. Lett.}\ }\textbf {\bibinfo
  {volume} {115}},\ \bibinfo {pages} {072501} (\bibinfo {year} {2015})},\
  \Eprint {http://arxiv.org/abs/1503.07514} {arXiv:1503.07514 [hep-th]}
  \BibitemShut {NoStop}%
\bibitem [{\citenamefont
  {Spali\'nski}(2018{\natexlab{a}})}]{Spalinski:2017mel}%
  \BibitemOpen
  \bibfield  {author} {\bibinfo {author} {\bibfnamefont {M.}~\bibnamefont
  {Spali\'nski}},\ }\href {\doibase 10.1016/j.physletb.2017.11.059} {\bibfield
  {journal} {\bibinfo  {journal} {Phys. Lett. B}\ }\textbf {\bibinfo {volume}
  {776}},\ \bibinfo {pages} {468} (\bibinfo {year} {2018}{\natexlab{a}})},\
  \Eprint {http://arxiv.org/abs/1708.01921} {arXiv:1708.01921 [hep-th]}
  \BibitemShut {NoStop}%
\bibitem [{\citenamefont {Strickland}\ \emph {et~al.}(2018)\citenamefont
  {Strickland}, \citenamefont {Noronha},\ and\ \citenamefont
  {Denicol}}]{Strickland:2017kux}%
  \BibitemOpen
  \bibfield  {author} {\bibinfo {author} {\bibfnamefont {M.}~\bibnamefont
  {Strickland}}, \bibinfo {author} {\bibfnamefont {J.}~\bibnamefont {Noronha}},
  \ and\ \bibinfo {author} {\bibfnamefont {G.}~\bibnamefont {Denicol}},\ }\href
  {\doibase 10.1103/PhysRevD.97.036020} {\bibfield  {journal} {\bibinfo
  {journal} {Phys. Rev. D}\ }\textbf {\bibinfo {volume} {97}},\ \bibinfo
  {pages} {036020} (\bibinfo {year} {2018})},\ \Eprint
  {http://arxiv.org/abs/1709.06644} {arXiv:1709.06644 [nucl-th]} \BibitemShut
  {NoStop}%
\bibitem [{\citenamefont {Strickland}(2018)}]{Strickland:2018ayk}%
  \BibitemOpen
  \bibfield  {author} {\bibinfo {author} {\bibfnamefont {M.}~\bibnamefont
  {Strickland}},\ }\href {\doibase 10.1007/JHEP12(2018)128} {\bibfield
  {journal} {\bibinfo  {journal} {JHEP}\ }\textbf {\bibinfo {volume} {12}},\
  \bibinfo {pages} {128} (\bibinfo {year} {2018})},\ \Eprint
  {http://arxiv.org/abs/1809.01200} {arXiv:1809.01200 [nucl-th]} \BibitemShut
  {NoStop}%
\bibitem [{\citenamefont
  {Spali\'nski}(2018{\natexlab{b}})}]{Spalinski:2018mqg}%
  \BibitemOpen
  \bibfield  {author} {\bibinfo {author} {\bibfnamefont {M.}~\bibnamefont
  {Spali\'nski}},\ }\href {\doibase 10.1016/j.physletb.2018.07.003} {\bibfield
  {journal} {\bibinfo  {journal} {Phys. Lett. B}\ }\textbf {\bibinfo {volume}
  {784}},\ \bibinfo {pages} {21} (\bibinfo {year} {2018}{\natexlab{b}})},\
  \Eprint {http://arxiv.org/abs/1805.11689} {arXiv:1805.11689 [hep-th]}
  \BibitemShut {NoStop}%
\bibitem [{\citenamefont {Giacalone}\ \emph {et~al.}(2019)\citenamefont
  {Giacalone}, \citenamefont {Mazeliauskas},\ and\ \citenamefont
  {Schlichting}}]{Giacalone:2019ldn}%
  \BibitemOpen
  \bibfield  {author} {\bibinfo {author} {\bibfnamefont {G.}~\bibnamefont
  {Giacalone}}, \bibinfo {author} {\bibfnamefont {A.}~\bibnamefont
  {Mazeliauskas}}, \ and\ \bibinfo {author} {\bibfnamefont {S.}~\bibnamefont
  {Schlichting}},\ }\href {\doibase 10.1103/PhysRevLett.123.262301} {\bibfield
  {journal} {\bibinfo  {journal} {Phys. Rev. Lett.}\ }\textbf {\bibinfo
  {volume} {123}},\ \bibinfo {pages} {262301} (\bibinfo {year} {2019})},\
  \Eprint {http://arxiv.org/abs/1908.02866} {arXiv:1908.02866 [hep-ph]}
  \BibitemShut {NoStop}%
\bibitem [{\citenamefont {Kurkela}\ \emph
  {et~al.}(2020{\natexlab{a}})\citenamefont {Kurkela}, \citenamefont {van~der
  Schee}, \citenamefont {Wiedemann},\ and\ \citenamefont
  {Wu}}]{Kurkela:2019set}%
  \BibitemOpen
  \bibfield  {author} {\bibinfo {author} {\bibfnamefont {A.}~\bibnamefont
  {Kurkela}}, \bibinfo {author} {\bibfnamefont {W.}~\bibnamefont {van~der
  Schee}}, \bibinfo {author} {\bibfnamefont {U.~A.}\ \bibnamefont {Wiedemann}},
  \ and\ \bibinfo {author} {\bibfnamefont {B.}~\bibnamefont {Wu}},\ }\href
  {\doibase 10.1103/PhysRevLett.124.102301} {\bibfield  {journal} {\bibinfo
  {journal} {Phys. Rev. Lett.}\ }\textbf {\bibinfo {volume} {124}},\ \bibinfo
  {pages} {102301} (\bibinfo {year} {2020}{\natexlab{a}})},\ \Eprint
  {http://arxiv.org/abs/1907.08101} {arXiv:1907.08101 [hep-ph]} \BibitemShut
  {NoStop}%
\bibitem [{\citenamefont {Denicol}\ and\ \citenamefont
  {Noronha}(2020)}]{Denicol:2019lio}%
  \BibitemOpen
  \bibfield  {author} {\bibinfo {author} {\bibfnamefont {G.~S.}\ \bibnamefont
  {Denicol}}\ and\ \bibinfo {author} {\bibfnamefont {J.}~\bibnamefont
  {Noronha}},\ }\href {\doibase 10.1103/PhysRevLett.124.152301} {\bibfield
  {journal} {\bibinfo  {journal} {Phys. Rev. Lett.}\ }\textbf {\bibinfo
  {volume} {124}},\ \bibinfo {pages} {152301} (\bibinfo {year} {2020})},\
  \Eprint {http://arxiv.org/abs/1908.09957} {arXiv:1908.09957 [nucl-th]}
  \BibitemShut {NoStop}%
\bibitem [{\citenamefont {Almaalol}\ \emph {et~al.}(2020)\citenamefont
  {Almaalol}, \citenamefont {Kurkela},\ and\ \citenamefont
  {Strickland}}]{Almaalol:2020rnu}%
  \BibitemOpen
  \bibfield  {author} {\bibinfo {author} {\bibfnamefont {D.}~\bibnamefont
  {Almaalol}}, \bibinfo {author} {\bibfnamefont {A.}~\bibnamefont {Kurkela}}, \
  and\ \bibinfo {author} {\bibfnamefont {M.}~\bibnamefont {Strickland}},\
  }\href {\doibase 10.1103/PhysRevLett.125.122302} {\bibfield  {journal}
  {\bibinfo  {journal} {Phys. Rev. Lett.}\ }\textbf {\bibinfo {volume} {125}},\
  \bibinfo {pages} {122302} (\bibinfo {year} {2020})},\ \Eprint
  {http://arxiv.org/abs/2004.05195} {arXiv:2004.05195 [hep-ph]} \BibitemShut
  {NoStop}%
\bibitem [{\citenamefont {Heller}\ \emph {et~al.}(2020)\citenamefont {Heller},
  \citenamefont {Jefferson}, \citenamefont {Spali\'nski},\ and\ \citenamefont
  {Svensson}}]{Heller:2020anv}%
  \BibitemOpen
  \bibfield  {author} {\bibinfo {author} {\bibfnamefont {M.~P.}\ \bibnamefont
  {Heller}}, \bibinfo {author} {\bibfnamefont {R.}~\bibnamefont {Jefferson}},
  \bibinfo {author} {\bibfnamefont {M.}~\bibnamefont {Spali\'nski}}, \ and\
  \bibinfo {author} {\bibfnamefont {V.}~\bibnamefont {Svensson}},\ }\href
  {\doibase 10.1103/PhysRevLett.125.132301} {\bibfield  {journal} {\bibinfo
  {journal} {Phys. Rev. Lett.}\ }\textbf {\bibinfo {volume} {125}},\ \bibinfo
  {pages} {132301} (\bibinfo {year} {2020})},\ \Eprint
  {http://arxiv.org/abs/2003.07368} {arXiv:2003.07368 [hep-th]} \BibitemShut
  {NoStop}%
\bibitem [{\citenamefont {Du}\ and\ \citenamefont
  {Schlichting}(2021)}]{Du:2020zqg}%
  \BibitemOpen
  \bibfield  {author} {\bibinfo {author} {\bibfnamefont {X.}~\bibnamefont
  {Du}}\ and\ \bibinfo {author} {\bibfnamefont {S.}~\bibnamefont
  {Schlichting}},\ }\href {\doibase 10.1103/PhysRevLett.127.122301} {\bibfield
  {journal} {\bibinfo  {journal} {Phys. Rev. Lett.}\ }\textbf {\bibinfo
  {volume} {127}},\ \bibinfo {pages} {122301} (\bibinfo {year} {2021})},\
  \Eprint {http://arxiv.org/abs/2012.09068} {arXiv:2012.09068 [hep-ph]}
  \BibitemShut {NoStop}%
\bibitem [{\citenamefont {Blaizot}\ and\ \citenamefont
  {Yan}(2021)}]{Blaizot:2021cdv}%
  \BibitemOpen
  \bibfield  {author} {\bibinfo {author} {\bibfnamefont {J.-P.}\ \bibnamefont
  {Blaizot}}\ and\ \bibinfo {author} {\bibfnamefont {L.}~\bibnamefont {Yan}},\
  }\href {\doibase 10.1103/PhysRevC.104.055201} {\bibfield  {journal} {\bibinfo
   {journal} {Phys. Rev. C}\ }\textbf {\bibinfo {volume} {104}},\ \bibinfo
  {pages} {055201} (\bibinfo {year} {2021})},\ \Eprint
  {http://arxiv.org/abs/2106.10508} {arXiv:2106.10508 [nucl-th]} \BibitemShut
  {NoStop}%
\bibitem [{\citenamefont {Chattopadhyay}\ \emph {et~al.}(2022)\citenamefont
  {Chattopadhyay}, \citenamefont {Jaiswal}, \citenamefont {Du}, \citenamefont
  {Heinz},\ and\ \citenamefont {Pal}}]{Chattopadhyay:2021ive}%
  \BibitemOpen
  \bibfield  {author} {\bibinfo {author} {\bibfnamefont {C.}~\bibnamefont
  {Chattopadhyay}}, \bibinfo {author} {\bibfnamefont {S.}~\bibnamefont
  {Jaiswal}}, \bibinfo {author} {\bibfnamefont {L.}~\bibnamefont {Du}},
  \bibinfo {author} {\bibfnamefont {U.}~\bibnamefont {Heinz}}, \ and\ \bibinfo
  {author} {\bibfnamefont {S.}~\bibnamefont {Pal}},\ }\href {\doibase
  10.1016/j.physletb.2021.136820} {\bibfield  {journal} {\bibinfo  {journal}
  {Phys. Lett. B}\ }\textbf {\bibinfo {volume} {824}},\ \bibinfo {pages}
  {136820} (\bibinfo {year} {2022})},\ \Eprint
  {http://arxiv.org/abs/2107.05500} {arXiv:2107.05500 [nucl-th]} \BibitemShut
  {NoStop}%
\bibitem [{\citenamefont {Du}\ \emph {et~al.}(2022)\citenamefont {Du},
  \citenamefont {Heller}, \citenamefont {Schlichting},\ and\ \citenamefont
  {Svensson}}]{Du:2022bel}%
  \BibitemOpen
  \bibfield  {author} {\bibinfo {author} {\bibfnamefont {X.}~\bibnamefont
  {Du}}, \bibinfo {author} {\bibfnamefont {M.~P.}\ \bibnamefont {Heller}},
  \bibinfo {author} {\bibfnamefont {S.}~\bibnamefont {Schlichting}}, \ and\
  \bibinfo {author} {\bibfnamefont {V.}~\bibnamefont {Svensson}},\ }\href
  {\doibase 10.1103/PhysRevD.106.014016} {\bibfield  {journal} {\bibinfo
  {journal} {Phys. Rev. D}\ }\textbf {\bibinfo {volume} {106}},\ \bibinfo
  {pages} {014016} (\bibinfo {year} {2022})},\ \Eprint
  {http://arxiv.org/abs/2203.16549} {arXiv:2203.16549 [hep-ph]} \BibitemShut
  {NoStop}%
\bibitem [{\citenamefont {Denicol}\ \emph {et~al.}(2014)\citenamefont
  {Denicol}, \citenamefont {Heinz}, \citenamefont {Martinez}, \citenamefont
  {Noronha},\ and\ \citenamefont {Strickland}}]{Denicol:2014tha}%
  \BibitemOpen
  \bibfield  {author} {\bibinfo {author} {\bibfnamefont {G.~S.}\ \bibnamefont
  {Denicol}}, \bibinfo {author} {\bibfnamefont {U.~W.}\ \bibnamefont {Heinz}},
  \bibinfo {author} {\bibfnamefont {M.}~\bibnamefont {Martinez}}, \bibinfo
  {author} {\bibfnamefont {J.}~\bibnamefont {Noronha}}, \ and\ \bibinfo
  {author} {\bibfnamefont {M.}~\bibnamefont {Strickland}},\ }\href {\doibase
  10.1103/PhysRevD.90.125026} {\bibfield  {journal} {\bibinfo  {journal} {Phys.
  Rev. D}\ }\textbf {\bibinfo {volume} {90}},\ \bibinfo {pages} {125026}
  (\bibinfo {year} {2014})},\ \Eprint {http://arxiv.org/abs/1408.7048}
  {arXiv:1408.7048 [hep-ph]} \BibitemShut {NoStop}%
\bibitem [{\citenamefont {Kurkela}\ \emph {et~al.}(2019)\citenamefont
  {Kurkela}, \citenamefont {Wiedemann},\ and\ \citenamefont
  {Wu}}]{Kurkela:2019kip}%
  \BibitemOpen
  \bibfield  {author} {\bibinfo {author} {\bibfnamefont {A.}~\bibnamefont
  {Kurkela}}, \bibinfo {author} {\bibfnamefont {U.~A.}\ \bibnamefont
  {Wiedemann}}, \ and\ \bibinfo {author} {\bibfnamefont {B.}~\bibnamefont
  {Wu}},\ }\href {\doibase 10.1140/epjc/s10052-019-7428-6} {\bibfield
  {journal} {\bibinfo  {journal} {Eur. Phys. J. C}\ }\textbf {\bibinfo {volume}
  {79}},\ \bibinfo {pages} {965} (\bibinfo {year} {2019})},\ \Eprint
  {http://arxiv.org/abs/1905.05139} {arXiv:1905.05139 [hep-ph]} \BibitemShut
  {NoStop}%
\bibitem [{\citenamefont {Kurkela}\ \emph
  {et~al.}(2020{\natexlab{b}})\citenamefont {Kurkela}, \citenamefont {Taghavi},
  \citenamefont {Wiedemann},\ and\ \citenamefont {Wu}}]{Kurkela:2020wwb}%
  \BibitemOpen
  \bibfield  {author} {\bibinfo {author} {\bibfnamefont {A.}~\bibnamefont
  {Kurkela}}, \bibinfo {author} {\bibfnamefont {S.~F.}\ \bibnamefont
  {Taghavi}}, \bibinfo {author} {\bibfnamefont {U.~A.}\ \bibnamefont
  {Wiedemann}}, \ and\ \bibinfo {author} {\bibfnamefont {B.}~\bibnamefont
  {Wu}},\ }\href {\doibase 10.1016/j.physletb.2020.135901} {\bibfield
  {journal} {\bibinfo  {journal} {Phys. Lett. B}\ }\textbf {\bibinfo {volume}
  {811}},\ \bibinfo {pages} {135901} (\bibinfo {year} {2020}{\natexlab{b}})},\
  \Eprint {http://arxiv.org/abs/2007.06851} {arXiv:2007.06851 [hep-ph]}
  \BibitemShut {NoStop}%
\bibitem [{\citenamefont {Ambrus}\ \emph {et~al.}(2022)\citenamefont {Ambrus},
  \citenamefont {Schlichting},\ and\ \citenamefont
  {Werthmann}}]{Ambrus:2022koq}%
  \BibitemOpen
  \bibfield  {author} {\bibinfo {author} {\bibfnamefont {V.~E.}\ \bibnamefont
  {Ambrus}}, \bibinfo {author} {\bibfnamefont {S.}~\bibnamefont {Schlichting}},
  \ and\ \bibinfo {author} {\bibfnamefont {C.}~\bibnamefont {Werthmann}},\
  }\href@noop {} {\  (\bibinfo {year} {2022})},\ \Eprint
  {http://arxiv.org/abs/2211.14379} {arXiv:2211.14379 [hep-ph]} \BibitemShut
  {NoStop}%
\bibitem [{\citenamefont {Ambru\cb{s}}\ \emph {et~al.}(2022)\citenamefont
  {Ambru\cb{s}}, \citenamefont {Schlichting},\ and\ \citenamefont
  {Werthmann}}]{Ambrus:2021fej}%
  \BibitemOpen
  \bibfield  {author} {\bibinfo {author} {\bibfnamefont {V.~E.}\ \bibnamefont
  {Ambru\cb{s}}}, \bibinfo {author} {\bibfnamefont {S.}~\bibnamefont
  {Schlichting}}, \ and\ \bibinfo {author} {\bibfnamefont {C.}~\bibnamefont
  {Werthmann}},\ }\href {\doibase 10.1103/PhysRevD.105.014031} {\bibfield
  {journal} {\bibinfo  {journal} {Phys. Rev. D}\ }\textbf {\bibinfo {volume}
  {105}},\ \bibinfo {pages} {014031} (\bibinfo {year} {2022})},\ \Eprint
  {http://arxiv.org/abs/2109.03290} {arXiv:2109.03290 [hep-ph]} \BibitemShut
  {NoStop}%
\bibitem [{\citenamefont {Borghini}\ \emph {et~al.}(2023)\citenamefont
  {Borghini}, \citenamefont {Borrell}, \citenamefont {Feld}, \citenamefont
  {Roch}, \citenamefont {Schlichting},\ and\ \citenamefont
  {Werthmann}}]{Borghini:2022iym}%
  \BibitemOpen
  \bibfield  {author} {\bibinfo {author} {\bibfnamefont {N.}~\bibnamefont
  {Borghini}}, \bibinfo {author} {\bibfnamefont {M.}~\bibnamefont {Borrell}},
  \bibinfo {author} {\bibfnamefont {N.}~\bibnamefont {Feld}}, \bibinfo {author}
  {\bibfnamefont {H.}~\bibnamefont {Roch}}, \bibinfo {author} {\bibfnamefont
  {S.}~\bibnamefont {Schlichting}}, \ and\ \bibinfo {author} {\bibfnamefont
  {C.}~\bibnamefont {Werthmann}},\ }\href {\doibase
  10.1103/PhysRevC.107.034905} {\bibfield  {journal} {\bibinfo  {journal}
  {Phys. Rev. C}\ }\textbf {\bibinfo {volume} {107}},\ \bibinfo {pages}
  {034905} (\bibinfo {year} {2023})},\ \Eprint
  {http://arxiv.org/abs/2209.01176} {arXiv:2209.01176 [hep-ph]} \BibitemShut
  {NoStop}%
\bibitem [{\citenamefont {Kapusta}(1980)}]{Kapusta:1980zz}%
  \BibitemOpen
  \bibfield  {author} {\bibinfo {author} {\bibfnamefont {J.~I.}\ \bibnamefont
  {Kapusta}},\ }\href {\doibase 10.1103/PhysRevC.21.1301} {\bibfield  {journal}
  {\bibinfo  {journal} {Phys. Rev. C}\ }\textbf {\bibinfo {volume} {21}},\
  \bibinfo {pages} {1301} (\bibinfo {year} {1980})}\BibitemShut {NoStop}%
\bibitem [{\citenamefont {Petersen}\ \emph {et~al.}(2009)\citenamefont
  {Petersen}, \citenamefont {Steinheimer}, \citenamefont {Bleicher},\ and\
  \citenamefont {Stocker}}]{Petersen:2009mz}%
  \BibitemOpen
  \bibfield  {author} {\bibinfo {author} {\bibfnamefont {H.}~\bibnamefont
  {Petersen}}, \bibinfo {author} {\bibfnamefont {J.}~\bibnamefont
  {Steinheimer}}, \bibinfo {author} {\bibfnamefont {M.}~\bibnamefont
  {Bleicher}}, \ and\ \bibinfo {author} {\bibfnamefont {H.}~\bibnamefont
  {Stocker}},\ }\href {\doibase 10.1088/0954-3899/36/5/055104} {\bibfield
  {journal} {\bibinfo  {journal} {J. Phys. G}\ }\textbf {\bibinfo {volume}
  {36}},\ \bibinfo {pages} {055104} (\bibinfo {year} {2009})},\ \Eprint
  {http://arxiv.org/abs/0902.4866} {arXiv:0902.4866 [nucl-th]} \BibitemShut
  {NoStop}%
\bibitem [{\citenamefont {Huovinen}\ and\ \citenamefont
  {Petersen}(2012)}]{Huovinen:2012is}%
  \BibitemOpen
  \bibfield  {author} {\bibinfo {author} {\bibfnamefont {P.}~\bibnamefont
  {Huovinen}}\ and\ \bibinfo {author} {\bibfnamefont {H.}~\bibnamefont
  {Petersen}},\ }\href {\doibase 10.1140/epja/i2012-12171-9} {\bibfield
  {journal} {\bibinfo  {journal} {Eur. Phys. J. A}\ }\textbf {\bibinfo {volume}
  {48}},\ \bibinfo {pages} {171} (\bibinfo {year} {2012})},\ \Eprint
  {http://arxiv.org/abs/1206.3371} {arXiv:1206.3371 [nucl-th]} \BibitemShut
  {NoStop}%
\bibitem [{\citenamefont {Karpenko}\ \emph {et~al.}(2014)\citenamefont
  {Karpenko}, \citenamefont {Huovinen},\ and\ \citenamefont
  {Bleicher}}]{Karpenko:2013wva}%
  \BibitemOpen
  \bibfield  {author} {\bibinfo {author} {\bibfnamefont {I.}~\bibnamefont
  {Karpenko}}, \bibinfo {author} {\bibfnamefont {P.}~\bibnamefont {Huovinen}},
  \ and\ \bibinfo {author} {\bibfnamefont {M.}~\bibnamefont {Bleicher}},\
  }\href {\doibase 10.1016/j.cpc.2014.07.010} {\bibfield  {journal} {\bibinfo
  {journal} {Comput. Phys. Commun.}\ }\textbf {\bibinfo {volume} {185}},\
  \bibinfo {pages} {3016} (\bibinfo {year} {2014})},\ \Eprint
  {http://arxiv.org/abs/1312.4160} {arXiv:1312.4160 [nucl-th]} \BibitemShut
  {NoStop}%
\bibitem [{\citenamefont {M\"uller}(1967)}]{Muller.1967}%
  \BibitemOpen
  \bibfield  {author} {\bibinfo {author} {\bibfnamefont {I.}~\bibnamefont
  {M\"uller}},\ }\href {\doibase 10.1007/BF01326412} {\bibfield  {journal}
  {\bibinfo  {journal} {Z. Phys.}\ }\textbf {\bibinfo {volume} {198}},\
  \bibinfo {pages} {329} (\bibinfo {year} {1967})}\BibitemShut {NoStop}%
\bibitem [{\citenamefont {Israel}\ and\ \citenamefont
  {Stewart}(1979)}]{Israel.1979}%
  \BibitemOpen
  \bibfield  {author} {\bibinfo {author} {\bibfnamefont {W.}~\bibnamefont
  {Israel}}\ and\ \bibinfo {author} {\bibfnamefont {J.~M.}\ \bibnamefont
  {Stewart}},\ }\href {\doibase 10.1016/0003-4916(79)90130-1} {\bibfield
  {journal} {\bibinfo  {journal} {Ann. Phys.}\ }\textbf {\bibinfo {volume}
  {118}},\ \bibinfo {pages} {341} (\bibinfo {year} {1979})}\BibitemShut
  {NoStop}%
\bibitem [{\citenamefont {Heiselberg}\ and\ \citenamefont
  {Levy}(1999)}]{Heiselberg:1998es}%
  \BibitemOpen
  \bibfield  {author} {\bibinfo {author} {\bibfnamefont {H.}~\bibnamefont
  {Heiselberg}}\ and\ \bibinfo {author} {\bibfnamefont {A.-M.}\ \bibnamefont
  {Levy}},\ }\href {\doibase 10.1103/PhysRevC.59.2716} {\bibfield  {journal}
  {\bibinfo  {journal} {Phys. Rev. C}\ }\textbf {\bibinfo {volume} {59}},\
  \bibinfo {pages} {2716} (\bibinfo {year} {1999})},\ \Eprint
  {http://arxiv.org/abs/nucl-th/9812034} {arXiv:nucl-th/9812034} \BibitemShut
  {NoStop}%
\bibitem [{\citenamefont {Borghini}\ and\ \citenamefont
  {Gombeaud}(2011)}]{Borghini:2010hy}%
  \BibitemOpen
  \bibfield  {author} {\bibinfo {author} {\bibfnamefont {N.}~\bibnamefont
  {Borghini}}\ and\ \bibinfo {author} {\bibfnamefont {C.}~\bibnamefont
  {Gombeaud}},\ }\href {\doibase 10.1140/epjc/s10052-011-1612-7} {\bibfield
  {journal} {\bibinfo  {journal} {Eur. Phys. J. C}\ }\textbf {\bibinfo {volume}
  {71}},\ \bibinfo {pages} {1612} (\bibinfo {year} {2011})},\ \Eprint
  {http://arxiv.org/abs/1012.0899} {arXiv:1012.0899 [nucl-th]} \BibitemShut
  {NoStop}%
\bibitem [{\citenamefont {Romatschke}(2018)}]{Romatschke:2018wgi}%
  \BibitemOpen
  \bibfield  {author} {\bibinfo {author} {\bibfnamefont {P.}~\bibnamefont
  {Romatschke}},\ }\href {\doibase 10.1140/epjc/s10052-018-6112-6} {\bibfield
  {journal} {\bibinfo  {journal} {Eur. Phys. J. C}\ }\textbf {\bibinfo {volume}
  {78}},\ \bibinfo {pages} {636} (\bibinfo {year} {2018})},\ \Eprint
  {http://arxiv.org/abs/1802.06804} {arXiv:1802.06804 [nucl-th]} \BibitemShut
  {NoStop}%
\bibitem [{\citenamefont {Kurkela}\ \emph {et~al.}(2018)\citenamefont
  {Kurkela}, \citenamefont {Wiedemann},\ and\ \citenamefont
  {Wu}}]{Kurkela:2018ygx}%
  \BibitemOpen
  \bibfield  {author} {\bibinfo {author} {\bibfnamefont {A.}~\bibnamefont
  {Kurkela}}, \bibinfo {author} {\bibfnamefont {U.~A.}\ \bibnamefont
  {Wiedemann}}, \ and\ \bibinfo {author} {\bibfnamefont {B.}~\bibnamefont
  {Wu}},\ }\href {\doibase 10.1016/j.physletb.2018.06.064} {\bibfield
  {journal} {\bibinfo  {journal} {Phys. Lett. B}\ }\textbf {\bibinfo {volume}
  {783}},\ \bibinfo {pages} {274} (\bibinfo {year} {2018})},\ \Eprint
  {http://arxiv.org/abs/1803.02072} {arXiv:1803.02072 [hep-ph]} \BibitemShut
  {NoStop}%
\bibitem [{\citenamefont {Borghini}\ \emph {et~al.}(2018)\citenamefont
  {Borghini}, \citenamefont {Feld},\ and\ \citenamefont
  {Kersting}}]{Borghini:2018xum}%
  \BibitemOpen
  \bibfield  {author} {\bibinfo {author} {\bibfnamefont {N.}~\bibnamefont
  {Borghini}}, \bibinfo {author} {\bibfnamefont {S.}~\bibnamefont {Feld}}, \
  and\ \bibinfo {author} {\bibfnamefont {N.}~\bibnamefont {Kersting}},\ }\href
  {\doibase 10.1140/epjc/s10052-018-6313-z} {\bibfield  {journal} {\bibinfo
  {journal} {Eur. Phys. J. C}\ }\textbf {\bibinfo {volume} {78}},\ \bibinfo
  {pages} {832} (\bibinfo {year} {2018})},\ \Eprint
  {http://arxiv.org/abs/1804.05729} {arXiv:1804.05729 [nucl-th]} \BibitemShut
  {NoStop}%
\bibitem [{\citenamefont {Kurkela}\ \emph {et~al.}(2021)\citenamefont
  {Kurkela}, \citenamefont {Mazeliauskas},\ and\ \citenamefont
  {T\"ornkvist}}]{Kurkela:2021ctp}%
  \BibitemOpen
  \bibfield  {author} {\bibinfo {author} {\bibfnamefont {A.}~\bibnamefont
  {Kurkela}}, \bibinfo {author} {\bibfnamefont {A.}~\bibnamefont
  {Mazeliauskas}}, \ and\ \bibinfo {author} {\bibfnamefont {R.}~\bibnamefont
  {T\"ornkvist}},\ }\href {\doibase 10.1007/JHEP11(2021)216} {\bibfield
  {journal} {\bibinfo  {journal} {JHEP}\ }\textbf {\bibinfo {volume} {11}},\
  \bibinfo {pages} {216} (\bibinfo {year} {2021})},\ \Eprint
  {http://arxiv.org/abs/2104.08179} {arXiv:2104.08179 [hep-ph]} \BibitemShut
  {NoStop}%
\bibitem [{\citenamefont {Bachmann}\ \emph {et~al.}(2022)\citenamefont
  {Bachmann}, \citenamefont {Borghini}, \citenamefont {Feld},\ and\
  \citenamefont {Roch}}]{Bachmann:2022cls}%
  \BibitemOpen
  \bibfield  {author} {\bibinfo {author} {\bibfnamefont {B.}~\bibnamefont
  {Bachmann}}, \bibinfo {author} {\bibfnamefont {N.}~\bibnamefont {Borghini}},
  \bibinfo {author} {\bibfnamefont {N.}~\bibnamefont {Feld}}, \ and\ \bibinfo
  {author} {\bibfnamefont {H.}~\bibnamefont {Roch}},\ }\href@noop {} {\
  (\bibinfo {year} {2022})},\ \bibinfo {note} {arXiv:2203.13306},\ \Eprint
  {http://arxiv.org/abs/2203.13306} {arXiv:2203.13306 [nucl-th]} \BibitemShut
  {NoStop}%
\bibitem [{\citenamefont {Bernhard}\ \emph {et~al.}(2019)\citenamefont
  {Bernhard}, \citenamefont {Moreland},\ and\ \citenamefont
  {Bass}}]{Bernhard:2019bmu}%
  \BibitemOpen
  \bibfield  {author} {\bibinfo {author} {\bibfnamefont {J.~E.}\ \bibnamefont
  {Bernhard}}, \bibinfo {author} {\bibfnamefont {J.~S.}\ \bibnamefont
  {Moreland}}, \ and\ \bibinfo {author} {\bibfnamefont {S.~A.}\ \bibnamefont
  {Bass}},\ }\href {\doibase 10.1038/s41567-019-0611-8} {\bibfield  {journal}
  {\bibinfo  {journal} {Nature Phys.}\ }\textbf {\bibinfo {volume} {15}},\
  \bibinfo {pages} {1113} (\bibinfo {year} {2019})}\BibitemShut {NoStop}%
\bibitem [{\citenamefont {Bjorken}(1983)}]{Bjorken:1982qr}%
  \BibitemOpen
  \bibfield  {author} {\bibinfo {author} {\bibfnamefont {J.~D.}\ \bibnamefont
  {Bjorken}},\ }\href {\doibase 10.1103/PhysRevD.27.140} {\bibfield  {journal}
  {\bibinfo  {journal} {Phys. Rev. D}\ }\textbf {\bibinfo {volume} {27}},\
  \bibinfo {pages} {140} (\bibinfo {year} {1983})}\BibitemShut {NoStop}%
\bibitem [{\citenamefont {Jankowski}\ \emph {et~al.}(2021)\citenamefont
  {Jankowski}, \citenamefont {Kamata}, \citenamefont {Martinez},\ and\
  \citenamefont {Spali\'nski}}]{Jankowski:2020itt}%
  \BibitemOpen
  \bibfield  {author} {\bibinfo {author} {\bibfnamefont {J.}~\bibnamefont
  {Jankowski}}, \bibinfo {author} {\bibfnamefont {S.}~\bibnamefont {Kamata}},
  \bibinfo {author} {\bibfnamefont {M.}~\bibnamefont {Martinez}}, \ and\
  \bibinfo {author} {\bibfnamefont {M.}~\bibnamefont {Spali\'nski}},\ }\href
  {\doibase 10.1103/PhysRevD.104.074012} {\bibfield  {journal} {\bibinfo
  {journal} {Phys. Rev. D}\ }\textbf {\bibinfo {volume} {104}},\ \bibinfo
  {pages} {074012} (\bibinfo {year} {2021})},\ \Eprint
  {http://arxiv.org/abs/2012.02184} {arXiv:2012.02184 [nucl-th]} \BibitemShut
  {NoStop}%
\bibitem [{\citenamefont {Niemi}\ and\ \citenamefont
  {Denicol}(2014)}]{Niemi:2014wta}%
  \BibitemOpen
  \bibfield  {author} {\bibinfo {author} {\bibfnamefont {H.}~\bibnamefont
  {Niemi}}\ and\ \bibinfo {author} {\bibfnamefont {G.~S.}\ \bibnamefont
  {Denicol}},\ }\href@noop {} {\  (\bibinfo {year} {2014})},\ \Eprint
  {http://arxiv.org/abs/1404.7327} {arXiv:1404.7327 [nucl-th]} \BibitemShut
  {NoStop}%
\bibitem [{\citenamefont {Denicol}\ \emph {et~al.}(2012)\citenamefont
  {Denicol}, \citenamefont {Niemi}, \citenamefont {Molnar},\ and\ \citenamefont
  {Rischke}}]{Denicol:2012cn}%
  \BibitemOpen
  \bibfield  {author} {\bibinfo {author} {\bibfnamefont {G.~S.}\ \bibnamefont
  {Denicol}}, \bibinfo {author} {\bibfnamefont {H.}~\bibnamefont {Niemi}},
  \bibinfo {author} {\bibfnamefont {E.}~\bibnamefont {Molnar}}, \ and\ \bibinfo
  {author} {\bibfnamefont {D.~H.}\ \bibnamefont {Rischke}},\ }\href {\doibase
  10.1103/PhysRevD.85.114047} {\bibfield  {journal} {\bibinfo  {journal} {Phys.
  Rev. D}\ }\textbf {\bibinfo {volume} {85}},\ \bibinfo {pages} {114047}
  (\bibinfo {year} {2012})},\ \bibinfo {note} {[Erratum: Phys.Rev.D 91, 039902
  (2015)]},\ \Eprint {http://arxiv.org/abs/1202.4551} {arXiv:1202.4551
  [nucl-th]} \BibitemShut {NoStop}%
\bibitem [{\citenamefont {Floerchinger}\ and\ \citenamefont
  {Wiedemann}(2011)}]{Floerchinger:2011pxy}%
  \BibitemOpen
  \bibfield  {author} {\bibinfo {author} {\bibfnamefont {S.}~\bibnamefont
  {Floerchinger}}\ and\ \bibinfo {author} {\bibfnamefont {U.~A.}\ \bibnamefont
  {Wiedemann}},\ }\href {\doibase 10.1007/JHEP11(2011)100} {\bibfield
  {journal} {\bibinfo  {journal} {JHEP}\ }\textbf {\bibinfo {volume} {11}},\
  \bibinfo {pages} {100} (\bibinfo {year} {2011})},\ \Eprint
  {http://arxiv.org/abs/1108.5535} {arXiv:1108.5535 [nucl-th]} \BibitemShut
  {NoStop}%
\bibitem [{\citenamefont {Martinez}\ and\ \citenamefont
  {Strickland}(2010)}]{Martinez:2010sc}%
  \BibitemOpen
  \bibfield  {author} {\bibinfo {author} {\bibfnamefont {M.}~\bibnamefont
  {Martinez}}\ and\ \bibinfo {author} {\bibfnamefont {M.}~\bibnamefont
  {Strickland}},\ }\href {\doibase 10.1016/j.nuclphysa.2010.08.011} {\bibfield
  {journal} {\bibinfo  {journal} {Nucl. Phys. A}\ }\textbf {\bibinfo {volume}
  {848}},\ \bibinfo {pages} {183} (\bibinfo {year} {2010})},\ \Eprint
  {http://arxiv.org/abs/1007.0889} {arXiv:1007.0889 [nucl-th]} \BibitemShut
  {NoStop}%
\bibitem [{\citenamefont {Florkowski}\ and\ \citenamefont
  {Ryblewski}(2011)}]{Florkowski:2010cf}%
  \BibitemOpen
  \bibfield  {author} {\bibinfo {author} {\bibfnamefont {W.}~\bibnamefont
  {Florkowski}}\ and\ \bibinfo {author} {\bibfnamefont {R.}~\bibnamefont
  {Ryblewski}},\ }\href {\doibase 10.1103/PhysRevC.83.034907} {\bibfield
  {journal} {\bibinfo  {journal} {Phys. Rev. C}\ }\textbf {\bibinfo {volume}
  {83}},\ \bibinfo {pages} {034907} (\bibinfo {year} {2011})},\ \Eprint
  {http://arxiv.org/abs/1007.0130} {arXiv:1007.0130 [nucl-th]} \BibitemShut
  {NoStop}%
\bibitem [{\citenamefont {Florkowski}\ \emph {et~al.}(2013)\citenamefont
  {Florkowski}, \citenamefont {Ryblewski},\ and\ \citenamefont
  {Strickland}}]{Florkowski:2013lya}%
  \BibitemOpen
  \bibfield  {author} {\bibinfo {author} {\bibfnamefont {W.}~\bibnamefont
  {Florkowski}}, \bibinfo {author} {\bibfnamefont {R.}~\bibnamefont
  {Ryblewski}}, \ and\ \bibinfo {author} {\bibfnamefont {M.}~\bibnamefont
  {Strickland}},\ }\href {\doibase 10.1103/PhysRevC.88.024903} {\bibfield
  {journal} {\bibinfo  {journal} {Phys. Rev. C}\ }\textbf {\bibinfo {volume}
  {88}},\ \bibinfo {pages} {024903} (\bibinfo {year} {2013})},\ \Eprint
  {http://arxiv.org/abs/1305.7234} {arXiv:1305.7234 [nucl-th]} \BibitemShut
  {NoStop}%
\bibitem [{\citenamefont {Martinez}\ \emph {et~al.}(2012)\citenamefont
  {Martinez}, \citenamefont {Ryblewski},\ and\ \citenamefont
  {Strickland}}]{Martinez:2012tu}%
  \BibitemOpen
  \bibfield  {author} {\bibinfo {author} {\bibfnamefont {M.}~\bibnamefont
  {Martinez}}, \bibinfo {author} {\bibfnamefont {R.}~\bibnamefont {Ryblewski}},
  \ and\ \bibinfo {author} {\bibfnamefont {M.}~\bibnamefont {Strickland}},\
  }\href {\doibase 10.1103/PhysRevC.85.064913} {\bibfield  {journal} {\bibinfo
  {journal} {Phys. Rev. C}\ }\textbf {\bibinfo {volume} {85}},\ \bibinfo
  {pages} {064913} (\bibinfo {year} {2012})},\ \Eprint
  {http://arxiv.org/abs/1204.1473} {arXiv:1204.1473 [nucl-th]} \BibitemShut
  {NoStop}%
\bibitem [{\citenamefont {McNelis}\ \emph {et~al.}(2021)\citenamefont
  {McNelis}, \citenamefont {Bazow},\ and\ \citenamefont
  {Heinz}}]{McNelis:2021zji}%
  \BibitemOpen
  \bibfield  {author} {\bibinfo {author} {\bibfnamefont {M.}~\bibnamefont
  {McNelis}}, \bibinfo {author} {\bibfnamefont {D.}~\bibnamefont {Bazow}}, \
  and\ \bibinfo {author} {\bibfnamefont {U.}~\bibnamefont {Heinz}},\ }\href
  {\doibase 10.1016/j.cpc.2021.108077} {\bibfield  {journal} {\bibinfo
  {journal} {Comput. Phys. Commun.}\ }\textbf {\bibinfo {volume} {267}},\
  \bibinfo {pages} {108077} (\bibinfo {year} {2021})},\ \Eprint
  {http://arxiv.org/abs/2101.02827} {arXiv:2101.02827 [nucl-th]} \BibitemShut
  {NoStop}%
\bibitem [{\citenamefont {Werner}(2007)}]{Werner:2007bf}%
  \BibitemOpen
  \bibfield  {author} {\bibinfo {author} {\bibfnamefont {K.}~\bibnamefont
  {Werner}},\ }\href {\doibase 10.1103/PhysRevLett.98.152301} {\bibfield
  {journal} {\bibinfo  {journal} {Phys. Rev. Lett.}\ }\textbf {\bibinfo
  {volume} {98}},\ \bibinfo {pages} {152301} (\bibinfo {year} {2007})},\
  \Eprint {http://arxiv.org/abs/0704.1270} {arXiv:0704.1270 [nucl-th]}
  \BibitemShut {NoStop}%
\bibitem [{\citenamefont {Kanakubo}\ \emph {et~al.}(2020)\citenamefont
  {Kanakubo}, \citenamefont {Tachibana},\ and\ \citenamefont
  {Hirano}}]{Kanakubo:2019ogh}%
  \BibitemOpen
  \bibfield  {author} {\bibinfo {author} {\bibfnamefont {Y.}~\bibnamefont
  {Kanakubo}}, \bibinfo {author} {\bibfnamefont {Y.}~\bibnamefont {Tachibana}},
  \ and\ \bibinfo {author} {\bibfnamefont {T.}~\bibnamefont {Hirano}},\ }\href
  {\doibase 10.1103/PhysRevC.101.024912} {\bibfield  {journal} {\bibinfo
  {journal} {Phys. Rev. C}\ }\textbf {\bibinfo {volume} {101}},\ \bibinfo
  {pages} {024912} (\bibinfo {year} {2020})},\ \Eprint
  {http://arxiv.org/abs/1910.10556} {arXiv:1910.10556 [nucl-th]} \BibitemShut
  {NoStop}%
\bibitem [{\citenamefont {Schenke}\ and\ \citenamefont
  {Venugopalan}(2014)}]{Schenke:2014zha}%
  \BibitemOpen
  \bibfield  {author} {\bibinfo {author} {\bibfnamefont {B.}~\bibnamefont
  {Schenke}}\ and\ \bibinfo {author} {\bibfnamefont {R.}~\bibnamefont
  {Venugopalan}},\ }\href {\doibase 10.1103/PhysRevLett.113.102301} {\bibfield
  {journal} {\bibinfo  {journal} {Phys. Rev. Lett.}\ }\textbf {\bibinfo
  {volume} {113}},\ \bibinfo {pages} {102301} (\bibinfo {year} {2014})},\
  \Eprint {http://arxiv.org/abs/1405.3605} {arXiv:1405.3605 [nucl-th]}
  \BibitemShut {NoStop}%
\bibitem [{\citenamefont {M\"antysaari}\ \emph {et~al.}(2022)\citenamefont
  {M\"antysaari}, \citenamefont {Schenke}, \citenamefont {Shen},\ and\
  \citenamefont {Zhao}}]{Mantysaari:2022ffw}%
  \BibitemOpen
  \bibfield  {author} {\bibinfo {author} {\bibfnamefont {H.}~\bibnamefont
  {M\"antysaari}}, \bibinfo {author} {\bibfnamefont {B.}~\bibnamefont
  {Schenke}}, \bibinfo {author} {\bibfnamefont {C.}~\bibnamefont {Shen}}, \
  and\ \bibinfo {author} {\bibfnamefont {W.}~\bibnamefont {Zhao}},\ }\href
  {\doibase 10.1016/j.physletb.2022.137348} {\bibfield  {journal} {\bibinfo
  {journal} {Phys. Lett. B}\ }\textbf {\bibinfo {volume} {833}},\ \bibinfo
  {pages} {137348} (\bibinfo {year} {2022})},\ \Eprint
  {http://arxiv.org/abs/2202.01998} {arXiv:2202.01998 [hep-ph]} \BibitemShut
  {NoStop}%
\bibitem [{\citenamefont {Demirci}\ \emph {et~al.}(2022)\citenamefont
  {Demirci}, \citenamefont {Lappi},\ and\ \citenamefont
  {Schlichting}}]{Demirci:2022wuy}%
  \BibitemOpen
  \bibfield  {author} {\bibinfo {author} {\bibfnamefont {S.}~\bibnamefont
  {Demirci}}, \bibinfo {author} {\bibfnamefont {T.}~\bibnamefont {Lappi}}, \
  and\ \bibinfo {author} {\bibfnamefont {S.}~\bibnamefont {Schlichting}},\
  }\href {\doibase 10.1103/PhysRevD.106.074025} {\bibfield  {journal} {\bibinfo
   {journal} {Phys. Rev. D}\ }\textbf {\bibinfo {volume} {106}},\ \bibinfo
  {pages} {074025} (\bibinfo {year} {2022})},\ \Eprint
  {http://arxiv.org/abs/2206.05207} {arXiv:2206.05207 [hep-ph]} \BibitemShut
  {NoStop}%
\bibitem [{\citenamefont {Brewer}\ \emph {et~al.}(2021)\citenamefont {Brewer},
  \citenamefont {Mazeliauskas},\ and\ \citenamefont {van~der
  Schee}}]{Brewer:2021kiv}%
  \BibitemOpen
  \bibfield  {author} {\bibinfo {author} {\bibfnamefont {J.}~\bibnamefont
  {Brewer}}, \bibinfo {author} {\bibfnamefont {A.}~\bibnamefont
  {Mazeliauskas}}, \ and\ \bibinfo {author} {\bibfnamefont {W.}~\bibnamefont
  {van~der Schee}},\ }in\ \href@noop {} {\emph {\bibinfo {booktitle}
  {{Opportunities of OO and pO collisions at the LHC}}}}\ (\bibinfo {year}
  {2021})\ \Eprint {http://arxiv.org/abs/2103.01939} {arXiv:2103.01939
  [hep-ph]} \BibitemShut {NoStop}%
\bibitem [{\citenamefont {Abelev}\ \emph {et~al.}(2022)\citenamefont {Abelev}
  \emph {et~al.}}]{ALICE:2022imr}%
  \BibitemOpen
  \bibfield  {author} {\bibinfo {author} {\bibfnamefont {B.~B.}\ \bibnamefont
  {Abelev}} \emph {et~al.} (\bibinfo {collaboration} {ALICE}),\ }\href@noop {}
  {\  (\bibinfo {year} {2022})},\ \Eprint {http://arxiv.org/abs/2204.10210}
  {arXiv:2204.10210 [nucl-ex]} \BibitemShut {NoStop}%
\bibitem [{\citenamefont {Abelev}\ \emph {et~al.}(2013)\citenamefont {Abelev}
  \emph {et~al.}}]{ALICE:2013rdo}%
  \BibitemOpen
  \bibfield  {author} {\bibinfo {author} {\bibfnamefont {B.~B.}\ \bibnamefont
  {Abelev}} \emph {et~al.} (\bibinfo {collaboration} {ALICE}),\ }\href
  {\doibase 10.1016/j.physletb.2013.10.054} {\bibfield  {journal} {\bibinfo
  {journal} {Phys. Lett. B}\ }\textbf {\bibinfo {volume} {727}},\ \bibinfo
  {pages} {371} (\bibinfo {year} {2013})},\ \Eprint
  {http://arxiv.org/abs/1307.1094} {arXiv:1307.1094 [nucl-ex]} \BibitemShut
  {NoStop}%
\bibitem [{\citenamefont {Nijs}\ and\ \citenamefont {van~der
  Schee}(2022)}]{Nijs:2021clz}%
  \BibitemOpen
  \bibfield  {author} {\bibinfo {author} {\bibfnamefont {G.}~\bibnamefont
  {Nijs}}\ and\ \bibinfo {author} {\bibfnamefont {W.}~\bibnamefont {van~der
  Schee}},\ }\href {\doibase 10.1103/PhysRevC.106.044903} {\bibfield  {journal}
  {\bibinfo  {journal} {Phys. Rev. C}\ }\textbf {\bibinfo {volume} {106}},\
  \bibinfo {pages} {044903} (\bibinfo {year} {2022})},\ \Eprint
  {http://arxiv.org/abs/2110.13153} {arXiv:2110.13153 [nucl-th]} \BibitemShut
  {NoStop}%
\end{thebibliography}%

\newpage

\clearpage

\newpage

\appendix

\begin{center}
 {\Large SUPPLEMENTAL MATERIAL}
\end{center}

\begin{figure*}
    \centering
    \includegraphics[width=.45\textwidth]{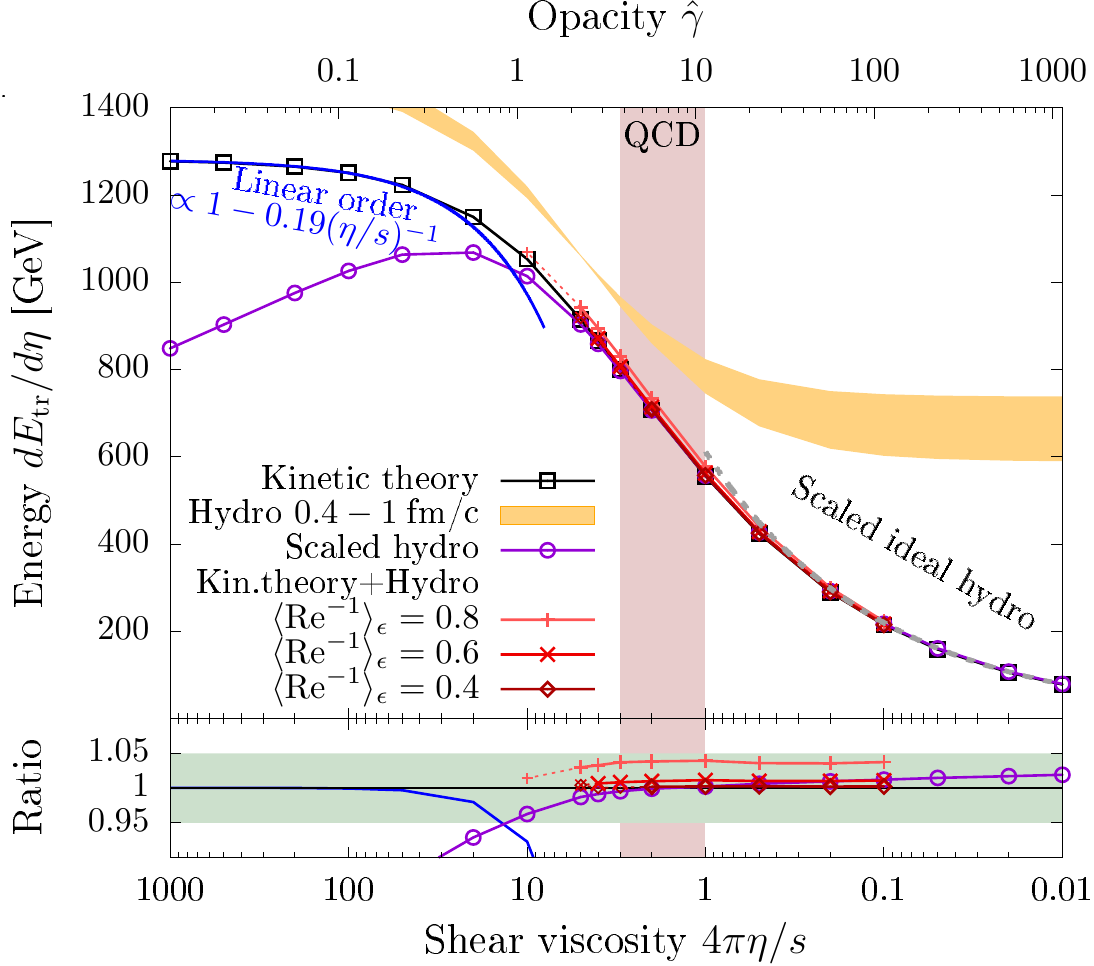}\includegraphics[width=.45\textwidth]{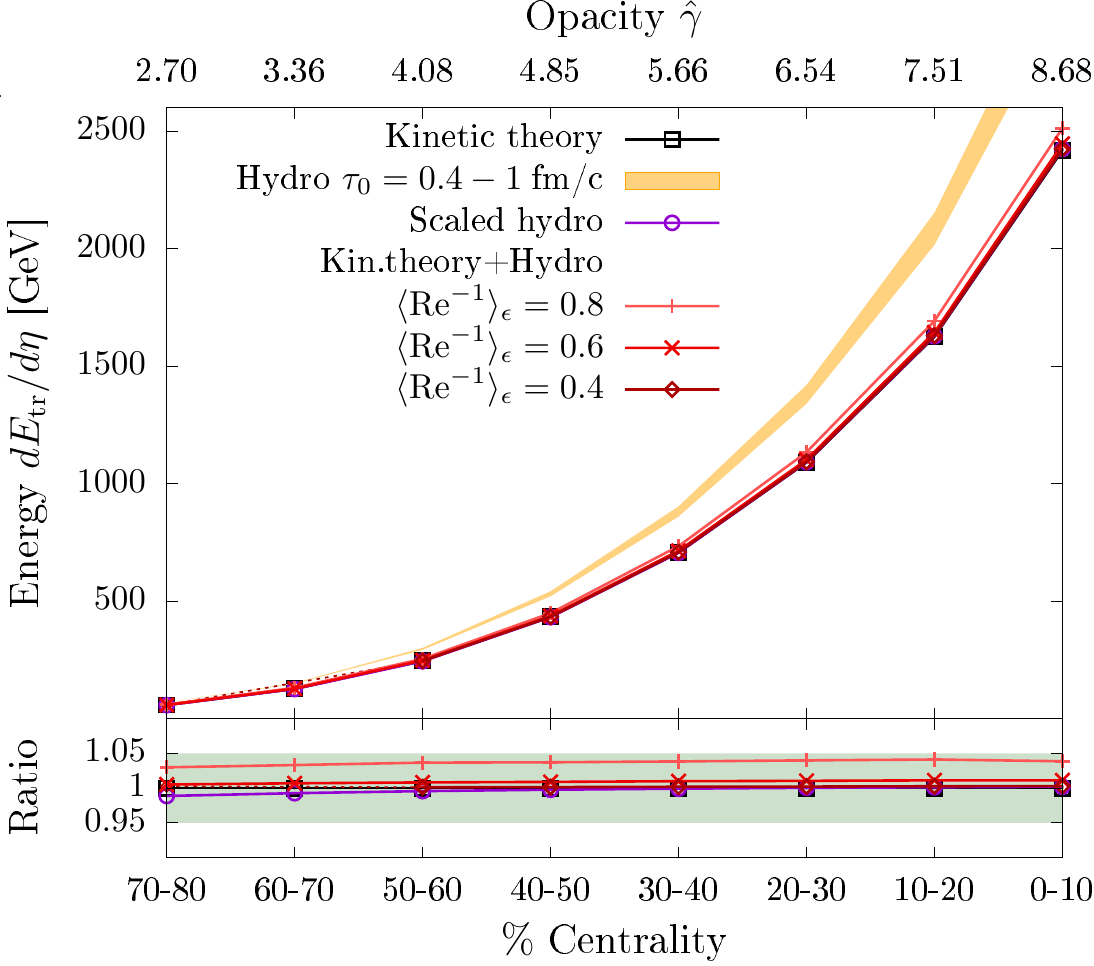}\\
    \includegraphics[width=.45\textwidth]{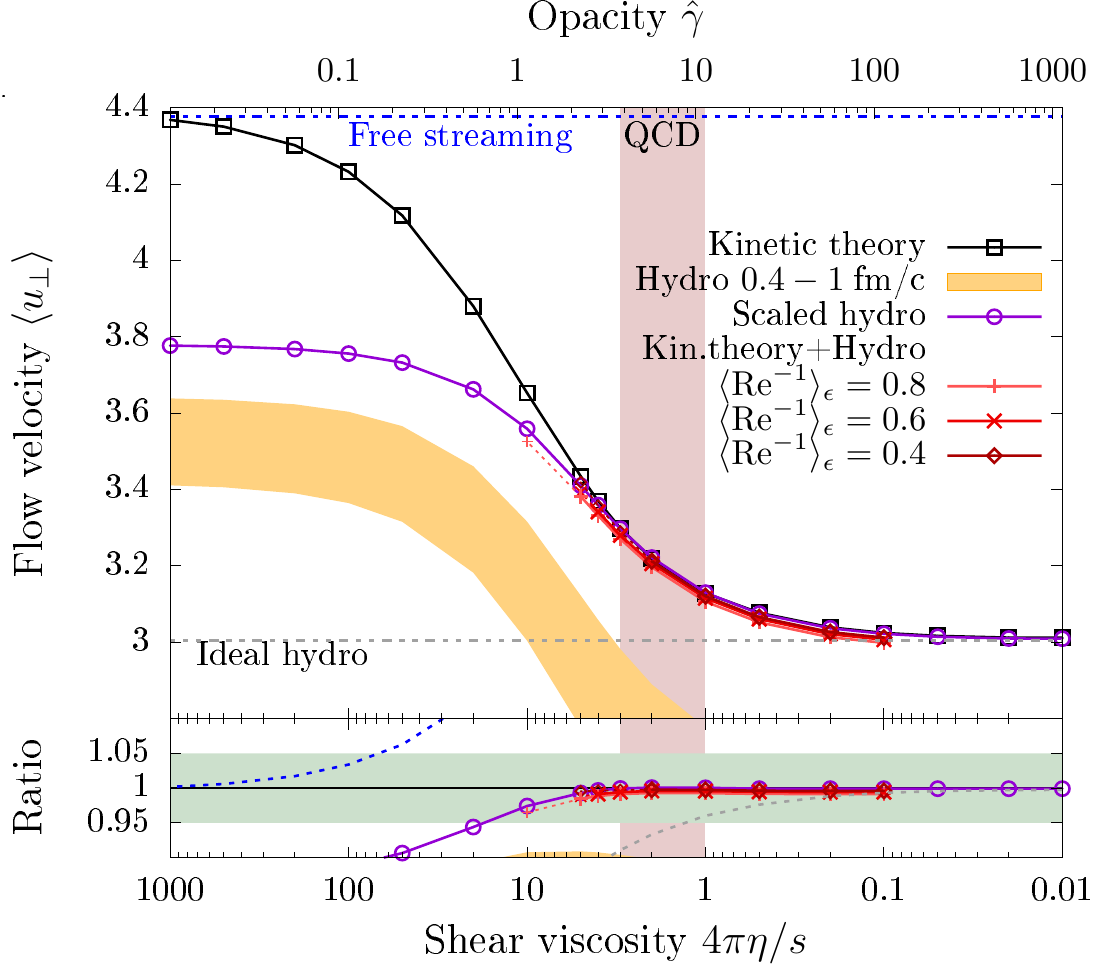}\includegraphics[width=.45\textwidth]{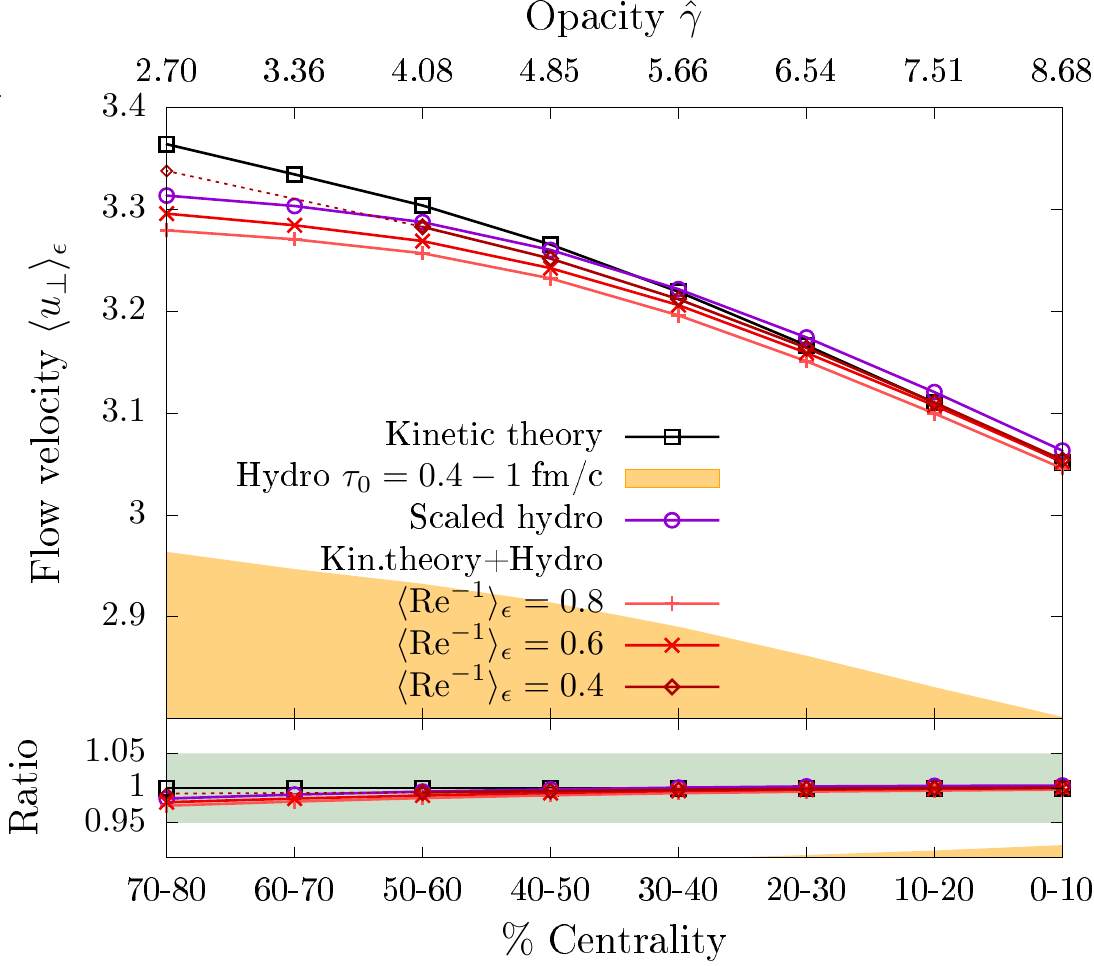}
    \caption{Variations of (top) transverse energy $dE_{\rm tr}/d\eta$ and (bottom) average transverse flow velocity $\eavg{u_\perp}$ (left) as a function of the shear-viscosity to entropy density ratio $\eta/s$ for $30-40\%$ $Pb+Pb$ collisions and (right) as a function of collision centrality for fixed $\eta/s=2/4\pi$. Simulations in kinetic theory (black squares) are compared to ideal (gray dashed) and viscous hydrodynamics (purple circles) as well as hybrid simulations (red pluses/crosses/diamonds) matching kinetic theory to hydrodynamics at different values of the average inverse Reynolds number $\text{Re}^{-1}$. Bands also show the results for naive hydrodynamics simulations with varying initialization times $\tau_{0}=0.4-1.0 {\rm fm}/c$. The blue curve in the upper left panel shows semi-analytic results from a leading order opacity expansion, while the lower left panel shows the free-streaming limit as a dashed blue curve.}
    \label{fig:master_totTxx+Tyy_uT}
\end{figure*}

\section{Additional Results}

In addition to the results for the elliptic flow response $\varepsilon_p$ presented in Fig.~\ref{fig:master_e_p}, we also tracked the transverse energy $dE_{\rm tr}/d\eta$ per unit rapidity and average transverse flow velocity $\eavg{u_\perp}$ in all model descriptions. These observables are defined as follows:
\begin{align}
    \frac{dE_{\rm tr}}{d\eta}&=\tau\int_\xT T^{xx}(\xT)+T^{yy}(\xT),\label{eq:dEdy_def}\\
    \eavg{u_\perp}&=\frac{\int_\xT \epsilon(\xT) [u_x^2(\xT)+u_y^2(\xT)]^{1/2}}{\int_\xT \epsilon(\xT)}.\label{eq:avguT_def}
\end{align}
Figure~\ref{fig:master_totTxx+Tyy_uT} shows results of the late time values ($\tau=4R$) of both observables for varied specific shear viscosity $\eta/s$ at fixed centrality (30-40\%) and varied centrality at a specific shear viscosity of $\eta/s=2/4\pi$.

In the case of varied opacity, results for $dE_{\rm tr}/d\eta$ are again compared to the estimate coming from the expansion in opacity, which is valid for opacities $\hat{\gamma} \lesssim 1$. In the case of $\eavg{u_\perp}$, kinetic theory results converge to the zero opacity ($\eta/s\to\infty$) free-streaming limit. In the large opacity limit ($\eta/s\to 0$), kinetic theory results converge to ideal hydrodynamics when it is initialized as described by Eq.~\eqref{eq:scaling}. It reproduces the large-opacity scaling law $\propto\hat{\gamma}^{-4/9}$ of kinetic theory.

Similar to the case of elliptic flow $\varepsilon_p$, for both varied specific shear viscosity and varied centrality, scaled viscous hydrodynamics is in excellent agreement with kinetic theory so long as transverse expansion sets in only after equilibration, i.e. for large opacities $\hat{\gamma} \gtrsim 3$. Hybrid simulations employing kinetic theory for pre-equilibrium also reach good accuracies and may slightly improve on scaled hydrodynamics at intermediate opacities $\hat{\gamma}\sim 3$. Overall, the results for $dE_{\rm tr}/d\eta$ and $\eavg{u_\perp}$ confirm that decreasing the value of the average inverse Reynolds number $\text{Re}^{-1}$ at which the dynamic description is switched to hydrodynamics will improve agreement with full kinetic theory.  Naive hydrodynamics initialized with the same energy density profile as kinetic theory at times $\tau_0=0.4 - 1\, \rm{fm}/c$ severely overestimates $dE_{\rm tr}/d\eta$ due to an early time increase in hydrodynamics and underestimates $\eavg{u_\perp}$ because of the absence of pre-flow, which again confirms the importance of pre-equilibrium dynamics.

\section{Scaling to match total energy}

The way that naive hydrodynamics was initialized in our comparisons differs from the common practice in the field. In phenomenological simulations, typically the initial energy density is scaled with a global factor that is chosen to match experimental results for the final state multiplicity. We investigated the effect of this by comparing also to a set of simulations where we mimic this common practice by matching the value of $dE_{\rm tr}/d\eta$ at late times $\tau=4R$. For varied specific shear viscosity $\eta/s$ at fixed centrality of $30-40\%$, all simulations were matched to kinetic theory at $4\pi\eta/s=2$, 
%i.e. the late time transverse energy was made to match the value 
when $dE_{\rm tr}/d\eta=707\,\mathrm{GeV}$. In the simulations scanning the centrality dependence at fixed $4\pi\eta/s=2$, a global scaling factor of the hydrodynamic initial condition was applied to make the late-time values of $dE_{\rm tr}/d\eta$ match with the one of kinetic theory, i.e.   $(2420, 1630, 1090, 707, 433, 246, 127, 58)\;$GeV  for the centrality classes $(0-10\%, 10-20\%, 20-30\%, 30-40\%, 40-50\%, 50-60\%, 60-70\%, 70-80\%)$.

\begin{figure*}
    \centering
    \includegraphics[width=.45\textwidth]{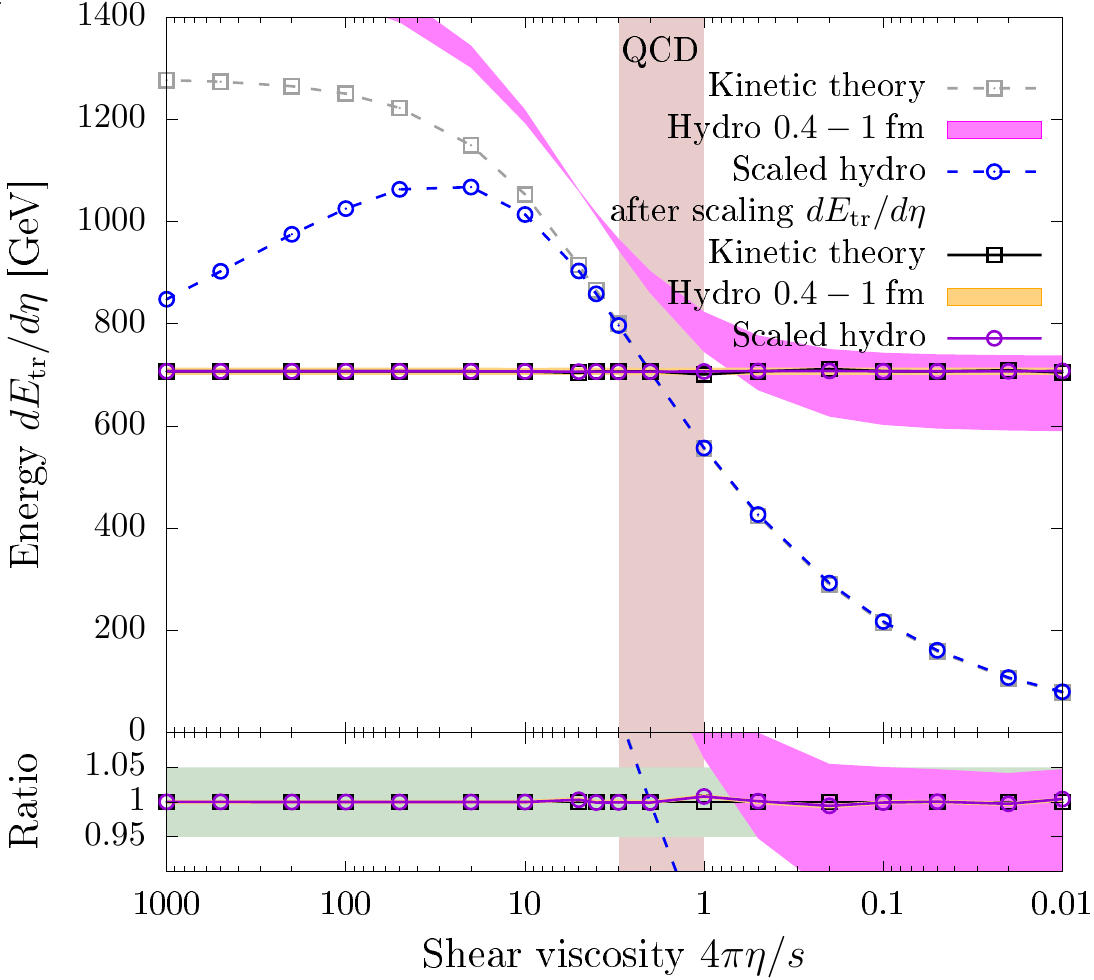}\includegraphics[width=.45\textwidth]{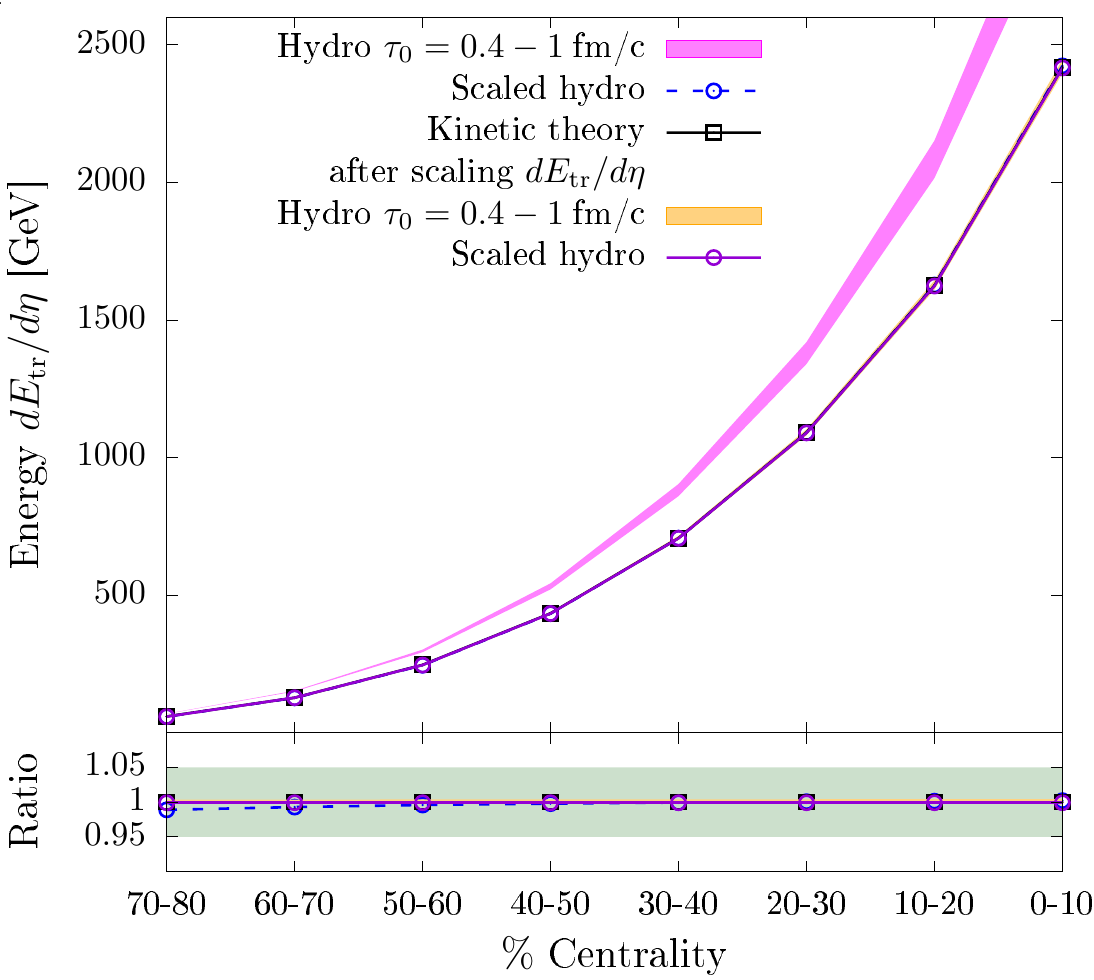}\\
    \includegraphics[width=.45\textwidth]{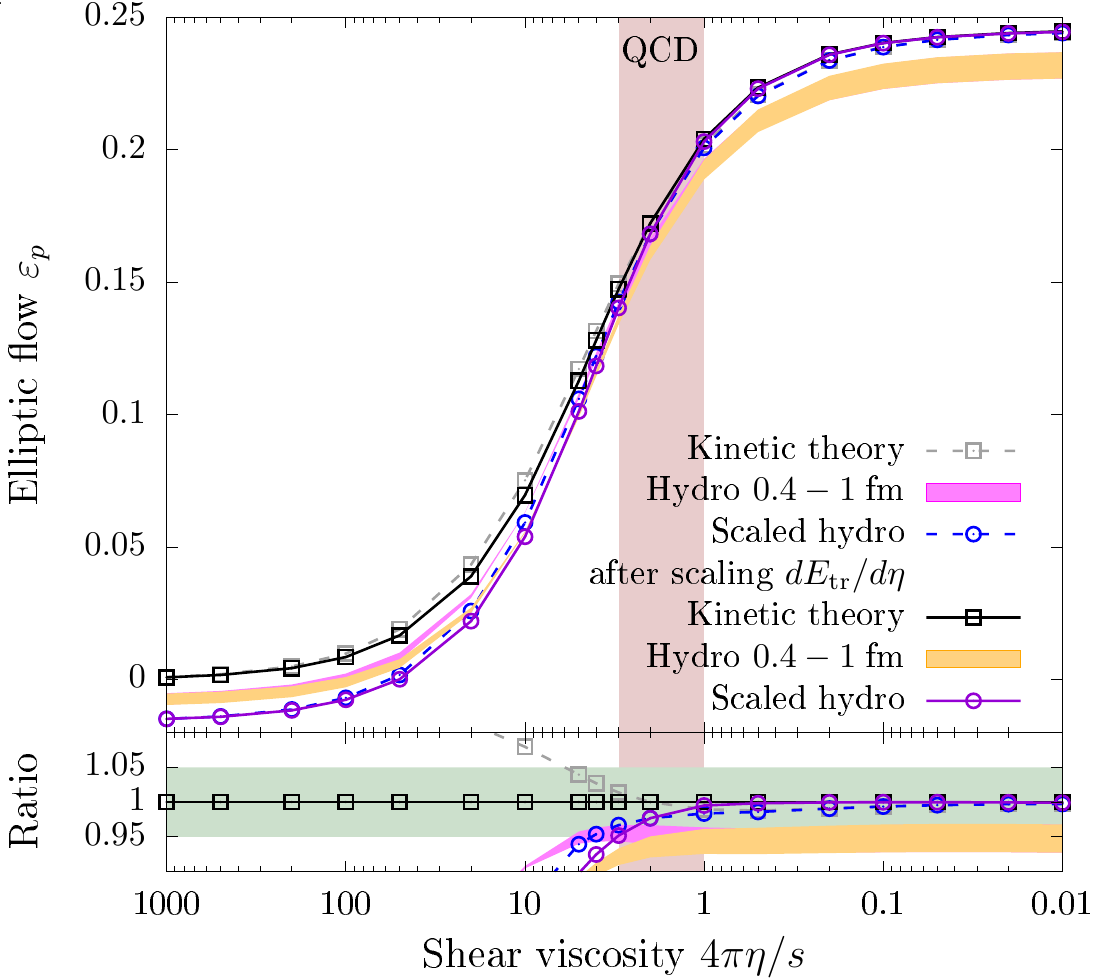}\includegraphics[width=.45\textwidth]{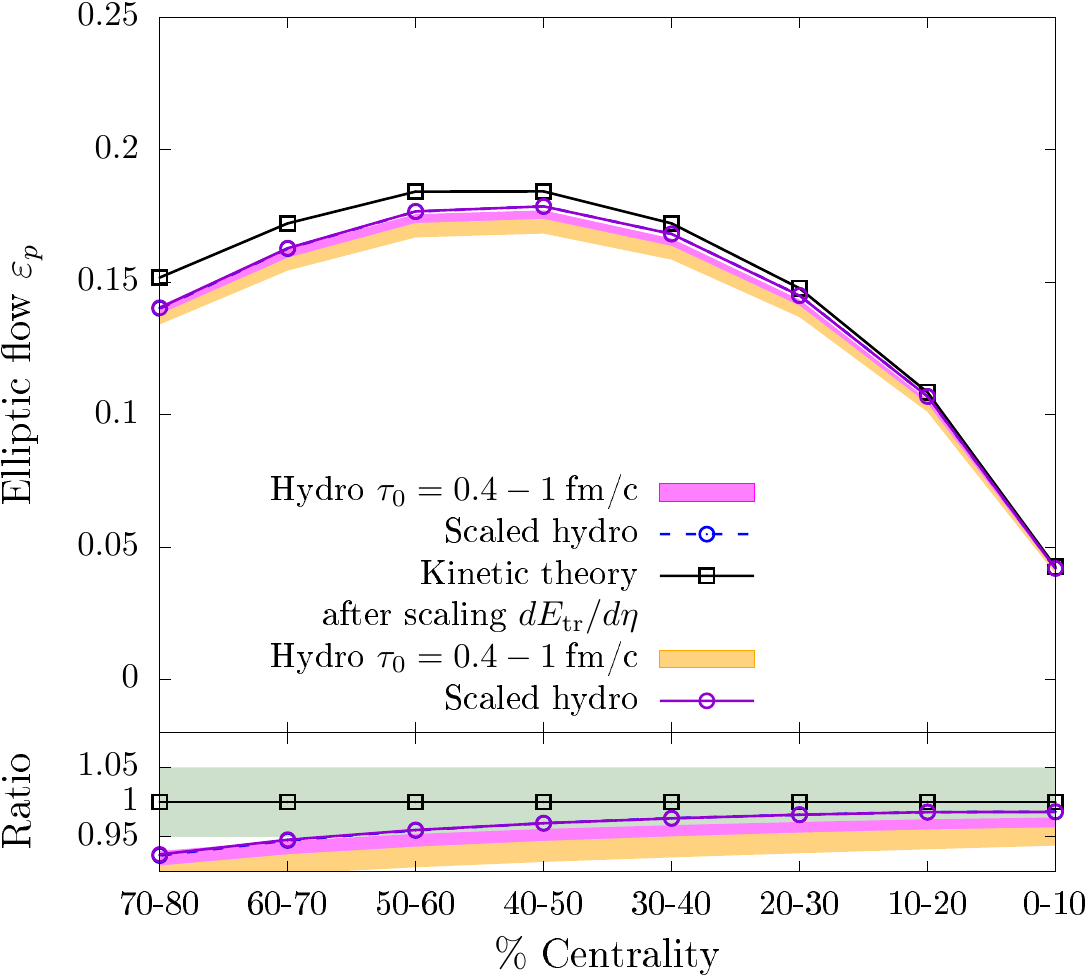}\\
    \includegraphics[width=.45\textwidth]{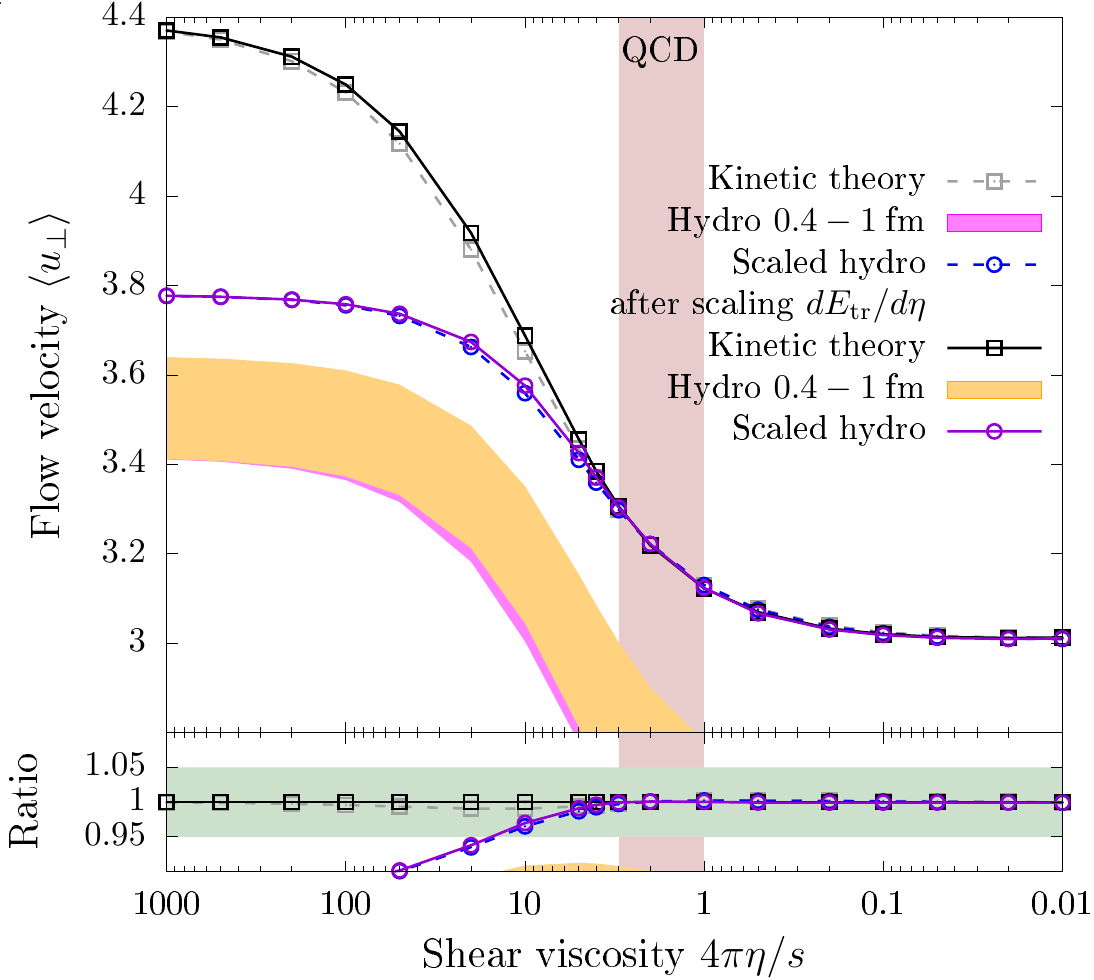}\includegraphics[width=.45\textwidth]{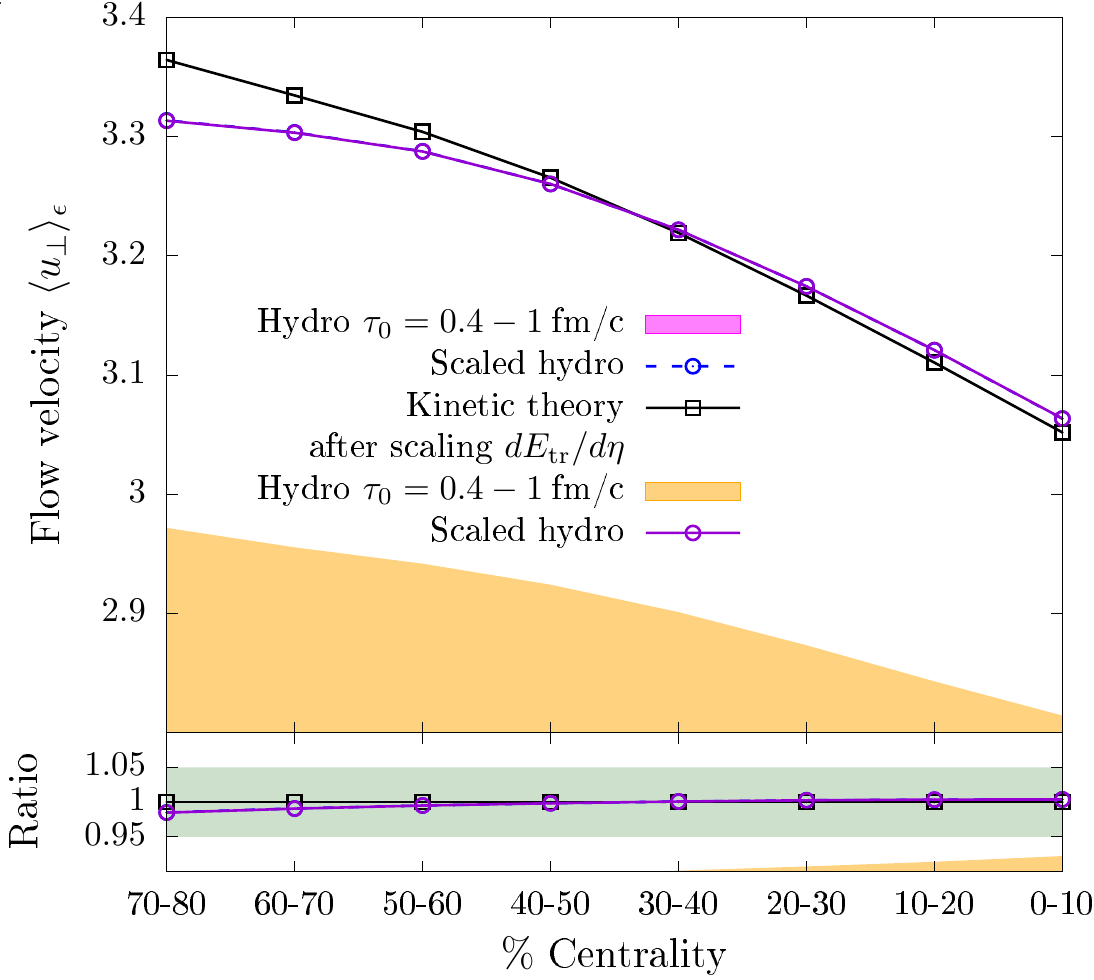}
    \caption{Variations of (middle) the elliptic flow $\varepsilon_p$ and (bottom) the average transverse flow velocity $\eavg{u_\perp}$ for scaled transverse energy as shown in the top row as a function of (left) $\eta/s$ for $30-40\%$ $Pb+Pb$ collisions and (right) of collision centrality for fixed $4\pi\eta/s=2$. Simulations in kinetic theory (black squares) are compared to viscous hydrodynamics (purple circles). Bands also show the results for naive hydrodynamics simulations with varying initialization times $\tau_{0}=0.4-1.0\ {\rm fm}/c$. Each plot also shows the same results when simulated in setups without the additional scaling of the initial condition obtained in kinetic theory (grey, dashed), scaled viscous hydrodynamics (blue, dashed) and naive hydrodynamics (magenta band).
    }
    \label{fig:master_constant_dEdy}
\end{figure*}

The results of these simulations for late time ($\tau=4R$) values of elliptic flow $\varepsilon_p$ and average transverse flow velocity $\eavg{u_\perp}$ are shown in Fig.~\ref{fig:master_constant_dEdy}. They are also compared to results without the global scaling factor to match final state energy, which are plotted with dashed lines and different colors. It is very apparent that the two observables are largely unaffected by this scaling scheme. The main effect of changing the initial state energy density in this way is that the opacity is increased for all simulations with specific shear viscosities below $4\pi\eta/s=2$ and decreased for those above this value. This slightly compresses the curves showing dependence on the specific shear viscosity in horizontal direction, but has very little effect on the vertical axes of the plots. For the centrality scan, the opacity was decreased in all plotted results, which very slightly decreases $\epsilon_p$ and increases $\eavg{u_\perp}$. However, most importantly the global scaling does not significantly improve the mismatch of kinetic theory and hydrodynamics at small opacities.

\section{Opacity estimates}

In our conclusions we make predictions on how close certain collision systems are to hydrodynamic behaviour based on an estimate of the value of the opacity parameter $\hat{\gamma}$ in each system. To compute $\hat{\gamma}$, we employed the estimates of the initial transverse energy $dE^{0}_\perp/d\eta$ and transverse radius $R$ compiled in Table~\ref{tab:opacity_estimates}. In accordance with the sources of the values taken to compute these estimates, the collisional energies of the listed collision systems are $7\;$TeV in the case of O+O and $5.02\;$TeV for the other systems.

\begin{table}
    \centering
    \begin{tabular}{c|c|c|c}
        System & $dE^{0}_{\bot}/d\eta$ [GeV] & $R$ [fm]  & $\hat{\gamma}$ \\ \hline
         p+p (min. bias) & 7.1 & 0.12 & 0.70\\
         p+Pb (min.bias) & 24 & 0.81 & 1.5 \\
         p+Pb (high mult.) & 230 & 0.81 & 2.7 \\ 
      %   p+Pb (0-5) & 39 & 0.8 & 1.7 \\ 
         O+O (70-80) & 13 & 0.88 & 1.4 \\
         O+O (30-40) & 55 & 1.13 & 2.2 \\
         O+O (0-5) & 140 & 1.61 & 3.1 \\
         Pb+Pb (70-80) & 85.1 & 2.16 & 2.70\\
         Pb+Pb (30-40) & 1280 & 2.78 & 5.66 \\
         Pb+Pb (0-5) & 5670 & 3.94 & 8.97
    \end{tabular}
    \caption{Estimated values of the initial transverse energy and system radius for various collision systems and the corresponding prediction of the value of the opacity parameter $\hat{\gamma}$ at a specific shear viscosity of $4\pi\eta/s=2$. 
    \label{tab:opacity_estimates}}
\end{table}

For an estimate of the radii, we used data from the initial state generator code that also produced the initial state for our main analysis~\cite{Borghini:2022iym}. In the same way, we obtained the initial energies for Pb+Pb collisions. In the other cases, the initial energy was computed according to an estimation formula given in~\cite{ALICE:2022imr}, which was modified to produce a mean estimate instead of a lower bound. Specifically, it states
\begin{align}
    \frac{dE^{(0)}_\perp}{d\eta}\approx \frac{1}{f_{\rm tot}}\sqrt{\langle m\rangle^2+\langle p_\perp \rangle^2}~\frac{dN_{\rm ch}}{d \eta}\;,
\end{align}
where $f_{\rm tot}=0.55$ is the ratio of charged to total particle number and $\langle m \rangle = 0.215\;$GeV is the average particle mass, while the mean transverse momentum $\langle p_\perp \rangle$ and the charged particle multiplicity $dN_{\rm ch}/d\eta$ have to be taken from data. 

 In practice, the values for $\langle p_\perp \rangle$ were taken as reported in~\cite{ALICE:2013rdo}. For the multiplicity, in the cases of minimum bias p+p and p+Pb we used data from the same paper that also proposed the estimation formula~\cite{ALICE:2022imr}, while for high multiplicity p+Pb we estimated $dN_{\rm ch}/d\eta\approx 150$, which according to results presented in~\cite{Dusling:2015gta} should roughly correspond to $0-0.1\%$ centrality. The multiplicities and mean transverse momenta for O+O were obtained from Trajectum simulation data~\cite{Nijs:2021clz}.

\end{document}